\documentstyle[12pt,aasms4]{article}

\slugcomment{submitted to the Astronomical Journal \today}

\lefthead{Grazian et al.}
\righthead{The Asiago-ESO/RASS QSO Survey}

\begin{document}

\title{The Asiago-ESO/RASS QSO Survey. I. \\
The Catalog and the Local QSO Luminosity Function \altaffilmark{*}}

\author{A. Grazian}
\affil{Astronomy Department, University of Padua, I-35122 Padua, Italy}

\author{S. Cristiani \altaffilmark{1}}
\affil{European Southern Observatory, ST European Coordinating
Facility, D-85748 Garching bei M\"unchen, Germany}

\author{V. D'Odorico}
\affil{SISSA, I-34013 Trieste, Italy} 

\author{A. Omizzolo}
\affil{Vatican Observatory}

\and

\author{A. Pizzella}
\affil{European Southern Observatory, Casilla 19001, Santiago 19,
Chile and Astronomy Department, University of Padua, I-35122 Padua, Italy}


\altaffiltext{1}{On leave from the Astronomy Department, University of Padua.}
\altaffilmark{*}{Based on material collected with the ESO-La Silla, Asiago, NOAO and
VATT telescopes}


\begin{abstract}
This paper presents the first results of a survey for bright quasars
($V < 14.5$ and $R<15.4$) covering the North Hemisphere at galactic
latitudes $|b|>30$.
The photometric database is derived from the GSC and USNO
catalogs. Quasars are identified on the basis of their X-ray emission
measured in the ROSAT All Sky Survey. The surface density of quasars
brighter than 15.5 mag turns out to be $10 \pm 2 \cdot 10^{-3} deg^{-2}$, 
about 3 times higher than that estimated by the PG survey. 
The quasar optical Luminosity Function (LF) 
at $0.04 < z \le 0.3$ is computed and shown to be consistent with a
Luminosity Dependent Luminosity Evolution of the type derived 
by La Franca and Cristiani (1997) in the range $0.3 < z \le 2.2$.
The predictions of semi-analytical models of
hierarchical structure formation agree remarkably well
with the present observations.
\end{abstract}




%
\section{Introduction}

Quasar surveys provide basic information for the understanding of a
number of astrophysical, cosmological and cosmogonical issues:
the formation and
evolution of galactic structures, the physics of the AGN phenomenon,
the UV and X-ray backgrounds.

The general behaviour of the QSO optical luminosity function (OLF)
is well established in the redshift
interval $0.3<z<2.2$ for which color techniques provide reliable selection
criteria (see \cite{HS:90,boyle92,HF94}, for a review of the subject).
A pure luminosity evolution (PLE) appears to reasonably describe tha data
in the interval $0.6<z<2.2$.  
In the range $0.3<z<0.6$ the OLF appears to be flatter than observed
at higher redshifts (\cite{pippa92}), requiring a luminosity dependent
luminosity evolution (LDLE, \cite{lf97} LC97). 
This departure from a PLE provides an interesting clue for the
physical interpretation of the QSO evolving population
(\cite{caval97}, LC97).
The present observational evidence, however, relies on a relatively
small number of objects: in the $0.3<z<0.6$ range for $M_B<-25$, $32$
QSOs are observed by LC97 instead of the
$19$ expected from the best-fit PLE model.  Analogous results by
K\"ohler et al. (1997) and Goldschmidt and Miller (1998) are likewise
based on very small samples.

To provide a statistically solid basis for the LDLE
pattern and to investigate whether such a trend
persists (and possibly becomes more evident) at redshifts lower than
0.3, we decided to carry out a new large-area survey of quasars at
bright optical fluxes.
A typical apparent magnitude for a $M_B = -24$ QSO at $z\sim 0.1$ is
in fact $B \sim 14.5$ and the surface density of these objects,
according to previous surveys (e.g. the BQS, \cite{SeG83}),
is expected to be less than $10^{-3}$ per sq.deg., requiring an efficient
selection criterion and the coverage of a significant fraction 
of the whole sky in order to collect a meaningful sample.

In Sect.2 we describe the photometric database from which optical
fluxes have been derived, in Sect.3 the criteria followed to select
the candidates, in Sect.4 the spectroscopic follow-up, in Sect.5 the
derived quasar counts and the optical luminosity function and in
Sect.6 a few consequences for model scenarios.

\section{The Photometric Database}

A number of photometric catalogues are available in the literature,
covering a substantial fraction of the celestial sphere down to the
optical magnitudes of interest for the present survey.
We have chosen the {\em USNO} (\cite{mone96}),  {\em GSC} (\cite{lask88})
and the {\em DSS} (\cite{http1}).

To test the accuracy of the photometric calibration of these
catalogues, we have used as a comparison in the Northern sky
the photometric standards of Landolt (1992).
In the Southern hemisphere we have used 446 standards
derived from the input catalog used to calibrate the photometric
material of the Homogeneous Bright Quasar Survey (HBQS, \cite{SC95}).
From both samples only relatively bright (12.5$\leq$$V$$\leq$16.0) and
not too red $(B-R)$$\leq$+1.0 stars have been chosen in order to match
the characteristics of the quasars searched for in this survey.

We finally defined three flux-limited samples adopting as photometric
reference: 
\begin{enumerate} 
\item {in the Northern hemisphere objects with $11.0 < V_{GSC} \le 14.5$ in
the GSC catalog. The relation between the $V_{GSC}$ band and
the corresponding Johnson $V$ turned out to be:
\begin{equation}
V_{GSC}=V-0.21
\end{equation}
with $\sigma_{V}=0.27$ $mag$. 7 ($4\%$) of the 183 Landolt stars 
used for this comparison were not found in the GSC catalog.
}
\item{in the Northern hemisphere objects from the USNO catalog with
$13.5 < R_{USNO} \le 15.4$.
The relation between the $R_{USNO}$ band and
the corresponding Johnson-Kron-Cousins $R$ turned out to be:
\begin{equation}
R_{USNO}=R_{JKC}+0.096
\end{equation}
with $\sigma_{R_{USNO}}=0.267$ $mag$.
No correlation between the residuals of $R_{USNO}-R_{JKC}$ and the
color $(B-R)_{JKC}$ has been found.  According to Monet et al. (1996),
the internal magnitude estimators for stars in the USNO catalog are
probably accurate to something like 0.15 magnitudes over the range
$R_{USNO}=12$ $mag$ to $R_{USNO}=19$ $mag$, but the systematic error
arising from the plate-to-plate differences is at least 0.25
magnitudes in the Northern hemisphere, consistent with our estimates.
}
\item {
in the Southern hemisphere we have derived
$B_J$ magnitudes from the Digitized Sky Survey (DSS). Small scans 
(``postage stamps'' of $2' \times 2'$) of each object
of interest and of 20-50 surrounding objects with known GSC $B_J$
magnitudes were extracted from the DSS. The magnitude of the object of 
interest was then calibrated against the GSC objects.
In this way a $\sigma_{B_J}$ of $0.10$ $mag$ was obtained in the interval
$12.0 < B_J <15.5$.
}
\end{enumerate} 

\section{The Selection of the Sample}

\subsection{The ROSAT All Sky Survey}
In order to pick out the bright quasars from the overwhelming number
of stars in the same magnitude range, an efficient selection criterion is
required. A convenient possibility is offered 
by the X-ray emission, which is a key signature of the AGN phenomenon.

The ROSAT All Sky Survey (hereafter RASS, \cite{vog92})
was carried out
during the period July 1990/February 1991 with the PSPC and has produced
a photometric database in the Soft-X (0.1-2.4 KeV). This shallow survey
covers almost the entire sky at a bright level ($10^{-13}$ erg 
cm$^{-2}$ sec$^{-1}$) and initially contained 60000 sources.
A more evolved reduction analysis, SASS-II, has produced
the RASS Bright Source Catalogue (RASS-BSC, \cite{vog99}),
a sample of
18811 X-ray sources at a limiting flux level of 0.05 $cps$\footnote{$cps$ is
the 'counts per second' of a source and is a measure of the flux, when a
precise X-soft spectrum is considered. For $F(\nu)\propto
\nu^{-1}$ 0.05 $cps$ correspond to $10^{-13}$ erg s$^{-1}$ cm$^{-2}$.
}
all over the sky.

The main constituents of the RASS catalogue are AGNs and peculiar
stars (CV, M stars, K stars, WDs, X-ray binaries and coronally active
stars), but there are also cluster of galaxies, BL Lacs, SN remnants,
neutron stars and normal galaxies (or starbursts).  A convenient way
to distinguish/isolate AGNs is the comparative analysis of their
soft-X/optical properties (\cite{hasing98}).

We have cross-correlated the RASS catalogue with the photometric
databases described in the previous section for sources at galactic
latitudes $|b|\ge 30$ deg and with a RASS-BSC exposure time $\ge 300$
s, i.e. $flux\ge$ 0.05 $cps$. 
Sources classified as extended in the RASS have been disregarded,
while no selection based on optical morphology was applied.
We have looked for optical objects in
the range $11.0 < V_{GSC} \le 14.5$ and $13.5 < R_{USNO} \le 15.4$
around the RASS sources, adopting a matching radius three times the
RMS positional uncertainty of each entry in the RASS catalog
(typically $3 \times 12''$). In this way very few (of the order of
$0.1 \%$) true identifications in the desired optical range are
missed. Misidentifications, i.e. ``X-ray quiet'' AGN in the desired
optical magnitude range falling by chance in the RASS error-box, are
also possible, but extremely unlikely, given the low surface density
of these bright AGN and are in any case irrelevant for the present
work, which aims at the definition of the local {\it optical} QSO LF.

The resulting catalogue covers 8164 square degrees in the North and
5855 square degrees in the South.

\subsection{ The $\alpha_{ox}$ distribution of quasars and the
Selection Criteria}

For each source the $\alpha_{ox}$ index was computed as
$\alpha_{ox}= -0.408 \log cps -0.163 R +3.65$
or
$\alpha_{ox}= -0.428 \log cps -0.171 V +3.84$
or 
$\alpha_{ox}= -0.483 \log cps -0.193 B_J +4.20$.

\begin{figure}
\plotone{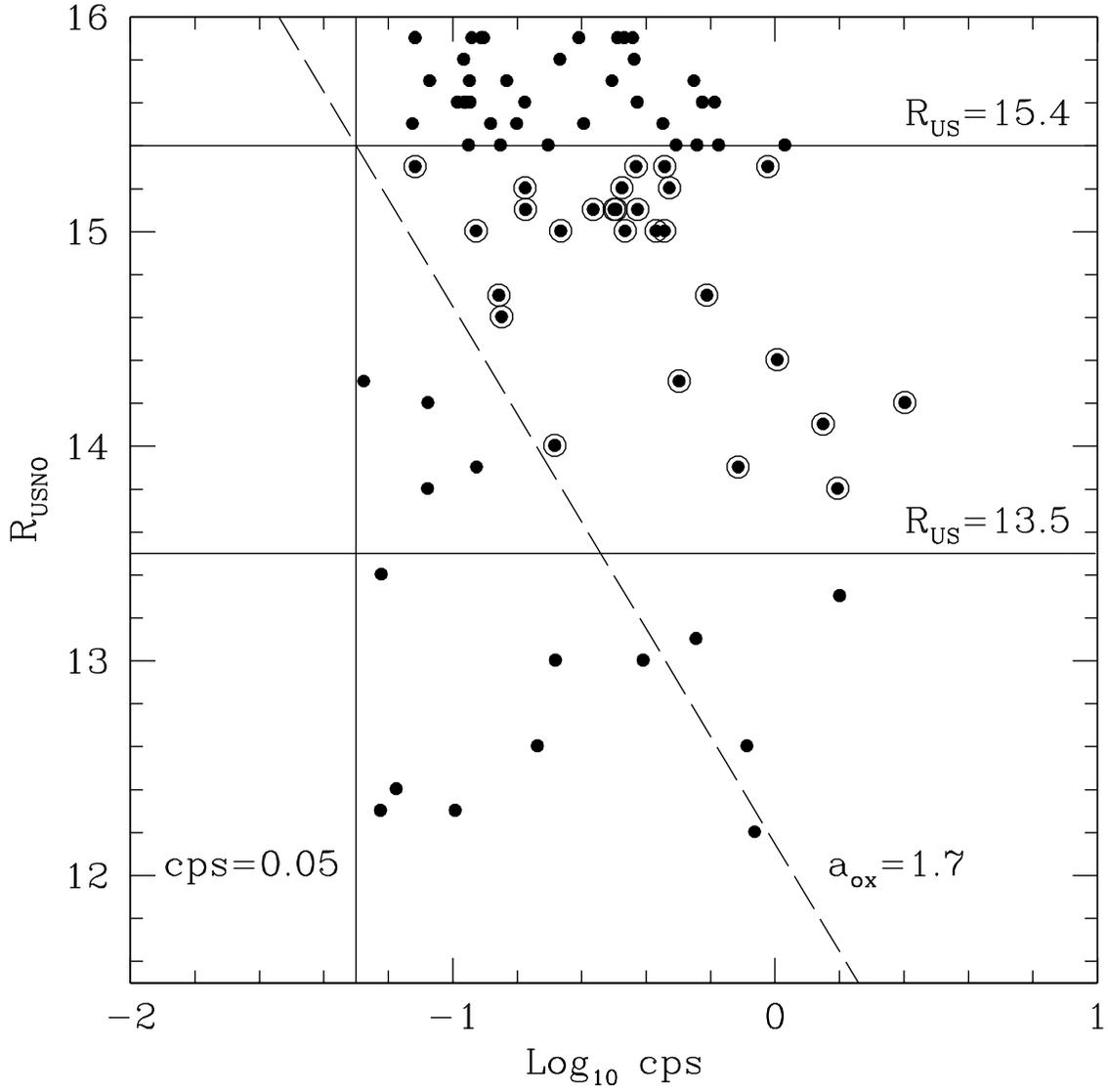}
\caption{The incompleteness in the present selection of quasar
candidates is due to objects not found in the RASS catalogue (with a
flux $\le 0.05$ cps, on the left of the vertical continuous line) and
to the ones with $\alpha_{ox}\ge 1.7$ (on the left of the dashed line).
\label{fig1}}
\end{figure}

To obtain an estimate of the intrinsic distribution of the
$\alpha_{ox}$ of quasars we have plotted the observed $B$, $V$ or $R$
magnitudes vs. the $\log cps$ for the $0.04< z \le 0.3$ quasars listed
in the $8^{th}$ edition
of the V\'eron catalog (VV98, 1998).
Fig.~1 shows the result for northern QSOs and $R$ magnitudes.
\placefigure{fig1}
In this diagram the locus $\alpha_{ox} = {\rm const}$ is represented
by a straight diagonal line. 
``X-ray quiet'' objects, i.e. with a flux$\le 0.05 cps$ to the left
of the vertical continuous line, are missed in the RASS.
If we assume that the intrinsic distribution of the $\alpha_{ox}$ is
not a function of the apparent luminosity (\cite{yuan:98}), selecting
objects with $\alpha_{ox} < \alpha_{max}$ (i.e. ``X-ray loud'') and
optically brighter than a convenient limit will
provide a sample with a degree of incompleteness that is not a
function of the apparent magnitude.  For example, in the case of the
USNO database the limit in $R$ turns out to be $R < (3.65
-\alpha_{max}-0.408 \log cps_{min}) /0.163 = 25.64 - 6.13
\alpha_{max}$. If we adopt (see Fig.~1) an $\alpha_{max}=1.7$ at magnitudes
brighter than $R=15.4$ only the objects with $\alpha_{ox} >
\alpha_{max}$ will be missed.

Tab.~1 lists the interval of optical magnitudes and the
corresponding limit on $\alpha_{ox}$ chosen for the USNO, GSC and DSS
sub-samples. The last column shows the degree of completeness
estimated on the basis of the fraction of quasars of the VV98 found with the
adopted criterion.
\placetable{tbl-1}
Examining the properties of the VV98 quasars in terms of the various
RASS parameters, we have found two further
empirical criteria, based on the hardness ratios $HR1$ and $HR2$
(\cite{vog99}), to increase the effectiveness of the selection
without affecting its completeness:
\begin{enumerate}
\item{
$-0.9\le HR1 \le 0.9 $, where $HR1$,
the hardness ratio 1 is defined as $(A-B)/(A+B)$, 
with the ROSAT-PSPC count
rates A in the hard band (0.5 $\div$ 2.0 keV) and B in the soft band (0.1
$\div$ 0.4 keV).}
\item{
$-0.6\leq HR2\leq +0.8$, where $HR2$,
the hardness ratio 2, is defined as $(C-D)/(C+D)$, 
with the ROSAT-PSPC count
rates C in the hard band (0.9 $\div$ 2.0 keV) and D in the soft band (0.5
$\div$ 0.9 keV).}
\end{enumerate}

\section{Spectroscopic Follow-up}

In the following we concentrate our discussion on the sample of
Northern objects, for which the follow-up spectroscopy is more
advanced.  The Southern sample will be described elsewhere. The list
of the quasar candidates and the results of the spectroscopy 
are reported in Tab.~2. 
It should be noted that Tab.~2 cannot be considered a list of
optical identifications of X-ray sources. The cross-correlation
procedure defined in the previous section aims at finding optical
objects in a desired magnitude range around X-ray sources. In some
cases an entry in Tab.~2 may exist even if the true identification of
the X-ray source is another (typically fainter) optical object. For
example, even if the optical counterpart of the RASS source
J013624.3+205712 is known to be the QSO 3C47.0, with $V\simeq 18.1$
and $z=0.425$, in Tab.~2 we list an object of $R_{USNO}=14.7$
which happens to fulfill the criteria of the cross-correlation.
\placetable{tbl-2a}
The follow-up observations of the QSO candidates have been carried out at
the 1.8m telescope in Asiago with a Boller and Chivens Spectrograph or
with the Asiago Faint Object Spectrograph and Camera (AFOSC),
at the 1.5m ESO, 1.5m Danish and NTT telescopes in La Silla
with a Boller and Chivens Spectrograph, DFOSC and EMMI respectively and
with the 90'' telescope at Kitt Peak.
The resolution of the spectra ranges between 10 and 30 \AA. 

The reduction process used the standard MIDAS facilities 
(Banse et al., 1983) available at the Padua Department of Astronomy and at 
ESO Garching.
The raw data were sky-subtracted and corrected for pixel-to-pixel sensitivity
variations by division with a suitably normalized exposure of the spectrum of
an incandescent source (flat-field). Wavelength calibration was carried out by
comparison with exposures of He-Ar, He, Ar and Ne lamps. Relative flux
calibration was carried out by observations of spectrophotometric
standard stars (Oke, 1990). 

The identification classes are: $AGN$ = emission-line object,
irrespective of the line width; $STAR$ = star; $GAL$ = galaxy; $BL$ = BL Lac 
object. Identifications as BL Lac or Galaxy have been taken from the
 NASA/IPAC Extragalactic Database (NED).
Uncertain identifications and redshifts are indicated with a
colon ($:$). 

In order to test the reliability of our selection, additional
candidates, selected with less restrictive criteria than those reported
in the previous section, were observed. They are
reported in Tab.~3.
The spectra of the AGNs found during the follow-up spectroscopy are shown in
Fig.~2.
\placetable{tbl-3}
\placefigure{fig2}
\section{First Results}

The spectroscopic observations of the Northern sample are still
incomplete: only 45$\%$ of the candidates have been identified.
Different areas of the sky, in particular different strips in right
ascension, have been observed down to different magnitude limits.
Tables 4-5 list the extension of the area covered with a complete
spectroscopic follow up as a function of the limiting magnitude.

In the following computations we have adopted for the Northern sample
the ``effective areas'' listed in Tabs.
~4-5, which take into account the
incompleteness factors estimated in the previous sections.
\placetable{tbl-4}
\placetable{tbl-5}
\subsection{Quasar Counts}

Fig.~3 shows the LogN-LogS relation of QSOs brighter than $M_B = -23$
mag
\footnote{In the present
paper the K-corrections are computed on the basis of the composite spectrum of
Cristiani and Vio (1990), galactic extinction is taken into account
according to Burstein and Heiles (1982). The values $q_0 = 0.5$, and
$H_0=50$~Km/s/Mpc are adopted throughout.}  
with $z \ge 0.04$, for the USNO and GSC subsamples
together. The LogN-LogS has been computed with the $\sum 1/Area_{max}$
method, a convenient approach when the various sub-areas have very different
magnitude limits. 

\placefigure{fig3}

The LogN-LogS relation found in the present survey is consistent with
a single power-law distribution with the slope $0.67$ reported by
K\"ohler et al. (1997) for QSOs with 0.07$\le z\le$2.2, with a
slightly higher normalization: if we fix $\beta = 0.67$ in a $\log N =
\beta B+k$ relation, we find $k=-12.15^{+0.17}_{-0.19}$.  If we
restrict our sample to $z>0.07$ we find $k=-12.2\pm 0.2$, in agreement
with the normalization, $k=-12.4$, of K\"ohler et
al. (1997).

A comparison with the Palomar Bright Quasar Survey (BQS, \cite{SeG83})
shows that at $B \sim 15.5$ the cumulative BQS counts are about a
factor 3 lower.  This confirms previous findings by Goldschmidt et
al.(1992), LC97, K\"ohler et al. (1997).  \placetable{tbl-counts}
 
\subsection{The QSO Luminosity Function at $0.04< z\le 0.3$}

A preliminary, cursory analysis of the QSO optical LF has been carried
out on the basis of the present sample. A more detailed discussion
of the complete spectroscopic database will be developed elsewhere 
(Grazian and Cristiani, in preparation).  To compute the LF at $0.04<
z\le 0.3$ we have used the generalized $1/V_{max}$ ``coherent''
estimator (\cite{avni:80}) in a slightly modified version which tries
to estimate in an unbiased manner the volume-luminosity space
``available'' to each object (cf. the method of Page \& Carrera, 1999)
and takes into account the evolution of the LF within the redshift
interval.  Errors were estimated from Poisson statistics
(\cite{gehrels:86}).  The data values of the LF at $0.04 < z \le 0.3$
are given in Tab.~\ref{tbl-LF}.  
\placetable{tbl-LF}
Figures 4a and 4b show the comparison of the newly derived QSO LF at
$0.04< z\le 0.3$ with data at higher redshift, up to $z=2.2$
(LC97), and with a PLE and an LDLE parameterization,
respectively.
The points in the range $0.04 < z \le 0.3$ are the result of the
present survey, 
the data in the other redshift ranges are derived from LC97.
No effort has been made in the derivation of the LF  at $0.04 < z \le
0.3$ to subtract off the luminosity of the host galaxy.
\placefigure{fig4}
The PLE parameterization is the global best fit to the QSO 
LF derived in the interval $0.3 < z < 2.2$  by LC97 (Model B), 
who found it to be inconsistent with
the data at  $0.3<z<0.6$ at a $3 \sigma$ level.
The present result confirms and strengthens the conclusion of LC97:
if we compare the prediction of the Model B PLE of LC97 in the range
$0.04< z\le 0.3$, a $\chi^2 \simeq 14$ for the 5 data points of
Tab.~\ref{tbl-LF} is derived, corresponding to a formal probability of
$1.9 \%$.  

Fig.~4b shows that an LDLE parameterization of the type of Model C of
LC97 can reproduce the data in a much more satisfactory way. 
The best agreement with the data from $z=0.04$ to $z=2$ is obtained,
assuming the functional form (LC97)
\begin{equation}
\label{eq:LDLE}
\Phi (M_B,z) = { {\Phi^\ast} \over { 10^{0.4[M_B-M_B^*(z)](\alpha + 1)} +
10^{0.4[M_B-M_B^*(z)](\beta + 1)} } } \\
\end{equation}
with
\begin{equation}
M_B^*(z) = M^*_B(z=2)-2.5 k \log [(1+z)/3]
\end{equation}
and
\begin{eqnarray}
for~M_B \leq M_B^*(z)&:&~ k = k_1 + k_2 [M_B-M_B^*(z)]e^{-z/{.40}} \\
\nonumber for~M_B > M_B^*(z)&:&~ k = k_1 
\end{eqnarray}
where $\alpha$ and $\beta$ correspond to the faint-end and bright-end
slopes of the optical LF, respectively and $M^*_B(z=2)$ is the
magnitude of the break in the double power-law shape of the LF at $z=2$.
The actual values adopted in the LDLE parameterization are:
$ {\Phi^\ast}= 9.8\times 10^{-7} mag^{-1} Mpc^{-3},
~M^*_B(z=2)=-26.3,
~k_1=3.33,
~k_2=0.37,
~\alpha = -1.45, 
~\beta = -3.76 $, providing a $\chi^2$ probability $\simeq 58 \%$ 
in the range $0.04 < z \le 0.3$ when compared to the 5 data points of
Tab.~\ref{tbl-LF}.
\section{Discussion}
Franceschini et al. (1994) and La Franca et al. (1995) have shown that
the QSO emission in the soft-X and at visible wavelengths scales
linearly.  A confirmation of the constant ratio $L_X/L_O$ and
independence both from $L_O$ and from $z$ comes from Yuan et al. (1998).
Boyle et al. (1994) derived a XLF comparable to the QSO OLF known at
that epoch with an evolutionary rate $L_X(z)=L_X(0)(1+z)^{3.25}$,
similar to the optical one. These works favor a scenario in which
essentially the same QSO population is observed, both in the soft-X
and in the optical. In this way the flattening of the OLF observed in
the present survey should be reflected in a flattening in the
corresponding soft-X LF. Indeed Miyaji et al. (1999) show that the
bright part of the 0.5-2 keV LF flattens with decreasing redshift from
a value $1+\beta \simeq 3.5$ at $0.4 < z < 0.8$ to $1+\beta \simeq
2.6$ at $0.015 < z < 0.2$, which is similar to what we observe in the
optical: $1+\beta \simeq 3.7$ at $0.6 < z < 1.0$ and $1+\beta \simeq
3.1$ at $0.04 < z \le 0.3$.

The decline of the space density of quasars from a peak around
redshift 2 to the present epoch has been modeled by several authors in 
the framework of hierarchical theories of structure formation
(\cite{cattaneo:99,HM:99,KH:99,MSD:99}). In particular Kauffmann and
Haehnelt (1999, KH99) have attempted to reproduce quantitatively the
evolution of the quasar number density incorporating a scheme for the
growth of massive black holes (MBHs) into semi-analytical models following
the evolution of galaxies in CDM-dominated scenarios.
Together with the decrease in the merging rates and in the amount of
gas available to fuel the MBHs, which are built-in features of the
semi-analytic models, KH99 assume an increase of the timescale for gas 
accretion in order to reproduce the steep decline in the number
density of quasars from $z \sim 2$ to $z=0$. Other authors have
followed similar recipes assuming a decreasing mass accretion
(\cite{HM:99}), or a decreasing efficiency of the accretion
(\cite{cattaneo:99}), or a delayed quasar activity with respect to the 
dynamical formation of the halos with a longer delay for smaller halos
(\cite{MSD:99}).  

As can be seen from Fig.~17 of KH99, the semi-analytical models have
difficulties in reproducing the steep decrease of
the QSO density at low redshift that is commonly measured (\cite{HS:90}).
The most promising scenario is a $\Lambda$CDM, in which the accretion
timescale, $t_{\rm acc}$, is assumed to vary in the same way as the
host galaxy dynamical time ($t_{\rm acc} \propto [0.7 +
0.3(1+z)^3]^{-1/2}$). This model is able to reproduce
the evolution of the galaxy LF and of the cold gas content of
galaxies, but is apparently predicting a too slow quasar decline.
The present data significantly reduce this disagreement in the
sense that the higher quasar space density measured in
our survey corresponds fairly well to the $\Lambda$CDM semi-analytical
predictions, as shown in Fig.~\ref{semi_anal}.
\placefigure{semi_anal}
\acknowledgments
It is a pleasure to thank A. Wicenec for his invaluable help with the
photometric database and F.La Franca, M. Haehnelt and G.Kauffmann for
enlightening discussions. We are indebted to Mira and Philippe V\'eron
whose indications helped us to improve significantly the accuracy of
the identifications.  We would also like to thank an anonymous referee
for several useful comments which improved the paper.  AG acknowledges
the generous hospitality of ESO during two visits to Garching.  This
work is based on photographic data obtained using The UK Schmidt
Telescope. The UK Schmidt Telescope was operated by the Royal
Observatory Edinburgh, with funding from the UK Science and
Engineering Research Council, until 1988 June, and thereafter by the
Anglo-Australian Observatory. Original plate material is copyright (c)
of the Royal Observatory Edinburgh and the Anglo-Australian
Observatory. The plates were processed into the present compressed
digital form with their permission. The Digitized Sky Survey was
produced at the Space Telescope Science Institute under US Government
grant NAG W-2166.  The NASA/IPAC Extragalactic Database (NED) is
operated by the Jet Propulsion Laboratory, California Institute of
Technology, under contract with the National Aeronautics and Space
Administration.
\clearpage
 
\begin{deluxetable}{lcccc}
\tablewidth{400pt}
\tablecaption{Selection Criteria and Completeness}
\tablehead{
\colhead{Sub-sample} &
\colhead{Magnitude Interval} &
\colhead{$\alpha_{ox}$} &
\colhead{Completeness} &
\colhead{Completeness} \\
\colhead{}          &
\colhead{}          &
\colhead{max}       &
\colhead{$N_{found}/N_{VV98}$} &
\colhead{ \% }
}
\startdata
USNO & $13.5 < R~ < 15.4   $ & 1.7 & 24/36 & 68 \nl
GSC  & $12.5 < V~ < 14.5   $ & 1.9 & 5/8 & 63 \nl
DSS  & $12.6 < B_J < 15.2 $ & 1.9 & 8/9 & 89 \nl
\label{tbl-1}
\enddata
\end{deluxetable}
\clearpage
\begin{deluxetable}{lrrrrrc}
\tablenum{2}
\tablewidth{0pt}
\tablecaption{THE USNO SAMPLE}
\tablehead{
\colhead{Name (1RXS)}      & \colhead{R.A.} 	&
\colhead{Declination}      & \colhead{$R_{USNO}$} & \colhead{$V_{GSC}$} &
\colhead{$z$}              & \colhead{Type}}
\startdata
 J000150.9+111705 & 00 01 50.6 &  +11 16 47.6 & 15.10 & & 0.158 & AGN \nl 
 J001031.3+105832 & 00 10 31.0 &  +10 58 29.6 & 13.60 & 14.66 & 0.089 & AGN \nl 
 J002337.1+044220 & 00 23 37.2 &	+4 42 22.3 & 14.30 & 15.13 & 0.081 & AGN \nl 
 J002811.6+310342 & 00 28 10.8 &  +31 03 47.8 & 15.30 & & 0.500 & AGN \nl 
 J002913.9+131605 & 00 29 13.7 &  +13 16 03.6 & 15.30 & 15.23 & 0.142 & AGN\nl 
 J004240.8+301742 & 00 42 41.6 &  +30 17 43.9 & 14.40 & &	   &	 \nl   
 J005154.8+172552 & 00 51 54.8 &  +17 25 58.4 & 15.30 & & 0.064 & AGN \nl 
 J013556.7+231605 & 01 35 58.5 &  +23 15 56.5 & 15.20 & &	   &	 \nl 
 J013624.3+205712 & 01 36 25.3 &  +20 56 49.6 & 14.70 & &  &  \nl
 J014239.9+000514 & 01 42 38.4 &	+0 05 14.9 & 15.30 & &	   &	 \nl 
 J015242.1+010040 & 01 52 41.9 &	+1 00 25.1 & 14.40 & 13.57 & 0.230 & GAL\nl 
 J015524.9+022818 & 01 55 24.9 &	+2 28 16.2 & 15.10 & &  & \nl 
 J015546.4+071902 & 01 55 46.4 &	+7 19 03.8 & 15.40 & &	   &	 \nl   
 J015935.1+104707 & 01 59 34.9 &  +10 46 46.7 & 15.20 & &	   &	 \nl   
 J075704.9+583245 & 07 57 06.2 &  +58 32 58.7 & 14.50 & 14.48 & & \nl 
 J080132.3+473618 & 08 01 32.0 &  +47 36 15.8 & 15.20 & & 0.158 & AGN \nl
 J080525.8+753424 & 08 05 23.7 &  +75 34 34.7 & 15.30 & 14.40 & & \nl 
 J080534.6+543132 & 08 05 34.8 &  +54 31 30.3 & 14.90 & &	   &	 \nl   
 J080649.5+751853 & 08 06 48.3 &  +75 18 30.7 & 15.10 & &	   &	 \nl 
 J080938.9+754851 & 08 09 39.8 &  +75 48 55.1 & 14.20 & & 0.094 & AGN \nl 
 J080949.2+521855 & 08 09 49.2 &  +52 18 58.2 & 14.50 & 14.84 & 0.138 & BL \nl 
 J081059.0+760245 & 08 10 58.6 &  +76 02 42.5 & 14.20 & 14.59 & 0.100 & AGN \nl 
 J081228.3+623627 & 08 12 28.2 &  +62 36 23.3 & 14.50 & 14.15 &	   &	 \nl   
 J083045.0+340527 & 08 30 45.4 &  +34 05 31.6 & 15.30 & & 0.063 & AGN \nl 
 J083121.0+483148 & 08 31 22.3 &  +48 32 13.9 & 15.10 & 15.13 &	   &	 \nl   
 J083821.6+483800 & 08 38 22.0 &  +48 38 01.7 & 14.80 & 14.63 &	   &	 \nl   
 J084255.9+292752 & 08 42 56.0 &  +29 27 26.2 & 15.40 & & 0.193 & GAL\nl
 J084445.2+765313 & 08 44 45.3 &  +76 53 09.3 & 13.90 & 15.72 & 0.131 & AGN \nl 
 J084658.3+704452 & 08 47 05.7 &  +70 44 41.1 & 15.20 & 14.62 &	   &	 \nl   
 J085343.5+574846 & 08 53 44.1 &  +57 48 41.3 & 15.00 & & 0.000 & STAR \nl
 J085358.8+770054 & 08 53 59.4 &  +77 00 54.6 & 15.40 & &	   &	 \nl   
 J085823.0+520533 & 08 58 24.2 &  +52 05 40.7 & 15.10 & &	   &	 \nl   
 J085902.0+484611 & 08 59 02.9 &  +48 46 09.0 & 14.30 & 14.99 & 0.083 & AGN \nl 
 J090020.1+503143 & 09 00 19.1 &  +50 31 40.5 & 15.10 & 14.91 &	   &	 \nl 
 J090038.4+411409 & 09 00 38.5 &  +41 13 55.9 & 15.10 & &	   &	 \nl 
 J090808.7+500912 & 09 08 08.8 &  +50 09 20.0 & 15.10 & &	   &	 \nl   
 J090950.6+184956 & 09 09 50.6 &  +18 49 47.6 & 15.30 & &	   &	 \nl   
 J091010.2+481317 & 09 10 10.0 &  +48 13 41.4 & 14.30 & & 0.118 & AGN \nl 
 J091254.7+793731 & 09 12 49.4 &  +79 37 51.8 & 14.70 & 15.59 &	   &	 \nl   
 J091552.3+090056 & 09 15 51.8 &	+9 00 50.9 & 14.50 & 12.99 & 0.000 & STAR \nl
 J091651.8+523829 & 09 16 52.0 &  +52 38 27.9 & 15.30 & & 0.190 & BL\nl
 J091904.6+732334 & 09 19 08.8 &  +73 23 59.2 & 15.30 & &	   &	 \nl   
 J091954.9+552120 & 09 19 55.3 &  +55 21 36.6 & 14.90 & & 0.123 & AGN \nl 
 J092246.4+512046 & 09 22 48.0 &  +51 20 45.9 & 14.40 & 14.80 &  & \nl 
 J092916.4+501344 & 09 29 15.7 &  +50 14 15.7 & 15.20 & 13.78 &  & \nl
 J093047.9+404446 & 09 30 47.8 &  +40 44 41.3 & 15.30 & &	   &	 \nl   
 J093355.6+141932 & 09 33 55.9 &  +14 19 19.9 & 15.40 & &	   &	 \nl   
 J093427.2+745123 & 09 34 28.4 &  +74 51 19.9 & 14.20 & &	   &	 \nl   
 J093701.0+010548 & 09 37 01.0 &	+1 05 43.0 & 13.60 & 13.87 & 0.051 & AGN \nl 
 J093942.8+560247 & 09 39 43.8 &  +56 02 30.6 & 14.10 & & 0.116 & AGN \nl 
 J094617.2+025505 & 09 46 16.9 &	+2 54 58.7 & 15.10 & &	   &	 \nl   
 J094653.0+132000 & 09 46 52.6 &  +13 19 53.6 & 15.20 & &  & \nl 
 J094713.2+762317 & 09 47 16.7 &  +76 23 28.2 & 15.00 & 14.79 &   & \nl   
 J095104.2+192531 & 09 51 03.5 &  +19 25 32.1 & 15.30 & &	   &	 \nl   
 J095406.7+212250 & 09 54 08.1 &  +21 22 25.5 & 14.90 & &	   &	 \nl 
 J095652.4+411524 & 09 56 52.3 &  +41 15 22.3 & 15.40 & 15.00 &   & \nl   
 J095708.1+243319 & 09 57 07.2 &  +24 33 15.6 & 14.50 & &	   &	 \nl   
 J100050.9+315555 & 10 00 52.1 &  +31 56 03.3 & 15.20 & &	   &	 \nl 
 J100121.5+555351 & 10 01 20.8 &  +55 53 52.8 & 14.80 & & 1.414 & AGN\nl
 J100335.1+444422 & 10 03 35.0 &  +44 44 39.6 & 15.20 & &	   &	 \nl   
 J100505.4+562426 & 10 05 06.1 &  +56 24 29.3 & 14.80 & &	   &	 \nl 
 J100659.7+673249 & 10 07 00.8 &  +67 32 46.8 & 14.00 & 14.69 &	   &	 \nl   
 J100851.6+541451 & 10 08 54.7 &  +54 14 45.9 & 15.40 & &	   &	 \nl
 J100947.3+523442 & 10 09 48.5 &  +52 34 51.2 & 15.20 & &       &     \nl
 J101238.4+101722 & 10 12 38.5 &  +10 17 18.8 & 14.80 & &	   &	 \nl   
 J101303.2+355131 & 10 13 03.2 &  +35 51 23.3 & 14.60 & 15.26 & 0.070 & AGN \nl 
 J101504.3+492604 & 10 15 04.1 &  +49 25 59.9 & 15.20 & & 0.200 & BL \nl
 J101624.0+333827 & 10 16 22.9 &  +33 38 17.0 & 14.10 & &	   &	 \nl   
 J101645.9+421024 & 10 16 45.1 &  +42 10 25.1 & 13.90 & & 0.054 & AGN \nl 
 J101702.4+390256 & 10 17 03.6 &  +39 02 49.4 & 14.30 & & 0.206 & GAL\nl 
 J101716.7+051145 & 10 17 16.8 &	+5 11 49.5 & 15.00 & &	   &	 \nl   
 J101906.8+231846 & 10 19 06.7 &  +23 18 37.0 & 15.20 & &	   &	 \nl 
 J102236.0+301753 & 10 22 37.4 &  +30 17 49.8 & 14.50 & 14.71 &	   &	 \nl 
 J102258.8+202252 & 10 22 58.2 &  +20 22 37.5 & 15.10 & & 0.129 & GAL \nl
 J102338.5+523356 & 10 23 39.6 &  +52 33 49.4 & 15.20 & &	   &	 \nl   
 J102531.2+514039 & 10 25 31.2 &  +51 40 34.7 & 13.60 & & 0.045 & AGN \nl 
 J102836.6+630255 & 10 28 37.2 &  +63 02 48.2 & 15.00 & &	   &	 \nl 
 J102915.4+572402 & 10 29 14.8 &  +57 23 53.1 & 15.30 & & 0.186 & AGN\nl
 J102946.7+401914 & 10 29 46.8 &  +40 19 13.6 & 14.70 & 15.07 & & \nl 
 J103134.2+284711 & 10 31 34.3 &  +28 47 00.6 & 15.20 & & 0.060 & AGN \nl 
 J103244.5+391331 & 10 32 44.2 &  +39 13 22.6 & 15.30 & &	   &	 \nl 
 J103422.3+605316 & 10 34 24.9 &  +60 53 11.5 & 15.40 & &	   &	 \nl   
 J103439.8+281755 & 10 34 39.9 &  +28 17 41.6 & 14.80 & &	   &	 \nl   
 J104043.6+330057 & 10 40 44.0 &  +33 00 59.5 & 14.90 & & 0.081 & AGN \nl 
 J104303.0+005423 & 10 43 02.5 &	+0 54 17.9 & 14.80 & &     &     \nl 
 J104346.6+223006 & 10 43 47.0 &  +22 29 57.4 & 14.60 & &	   &	 \nl 
 J104427.6+271813 & 10 44 27.7 &  +27 18 05.4 & 14.00 & &	   &	 \nl   
 J104427.6+271813 & 10 44 27.7 &  +27 18 05.4 & 14.00 & &	   &	 \nl   
 J104819.1+521837 & 10 48 18.0 &  +52 18 30.3 & 14.70 & &	   &	 \nl 
 J104926.1+245134 & 10 49 25.5 &  +24 51 23.0 & 14.70 & &	   &	 \nl 
 J105037.1+801204 & 10 50 35.6 &  +80 11 50.7 & 14.90 & 14.85 &	   &	 \nl 
 J105124.8+382053 & 10 51 24.5 &  +38 20 46.7 & 15.20 & &	   &	 \nl   
 J105143.8+335936 & 10 51 43.8 &  +33 59 26.5 & 15.10 & & 0.167 & AGN \nl 
 J105151.0+213739 & 10 51 51.0 &  +21 37 25.9 & 14.20 & &	   &	 \nl   
 J105214.2+055514 & 10 52 15.3 &	+5 55 08.2 & 15.00 & 13.96 &	   &	 \nl   
 J105355.0+661209 & 10 53 55.7 &  +66 12 01.8 & 15.30 & &	   &	 \nl   
 J105444.4+483145 & 10 54 44.7 &  +48 31 38.8 & 15.40 & & 0.286 & AGN \nl 
 J105519.1+402739 & 10 55 19.5 &  +40 27 16.6 & 15.20 & & 0.120 & AGN \nl 
 J105837.5+562816 & 10 58 37.7 &  +56 28 11.4 & 14.10 & & 0.144 & BL \nl
 J110237.0+724633 & 11 02 38.3 &  +72 46 20.7 & 15.10 & &  & \nl 
 J110412.4+765859 & 11 04 13.8 &  +76 58 58.2 & 15.40 & & 0.313 & AGN \nl 
 J110537.4+585128 & 11 05 37.6 &  +58 51 20.7 & 15.20 & &	   &	 \nl 
 J110748.8+710538 & 11 07 52.2 &  +71 06 01.4 & 14.40 & 14.75 &	   &	 \nl   
 J110831.9+695129 & 11 08 26.9 &  +69 51 41.7 & 15.00 & &	   &	 \nl   
 J111011.4+011333 & 11 10 12.1 &	+1 13 27.2 & 14.50 & 15.06 &	   &	 \nl   
 J111422.6+582318 & 11 14 21.9 &  +58 23 19.0 & 14.70 & & 0.206 & GAL\nl 
 J111830.0+402557 & 11 18 30.4 &  +40 25 54.5 & 14.40 & & 0.154 & AGN \nl 
 J111907.1+413018 & 11 19 07.6 &  +41 30 03.0 & 15.40 & &	   &	 \nl 
 J112034.0+100821 & 11 20 34.2 &  +10 08 04.5 & 15.20 & &	   &	 \nl   
 J112147.3+114420 & 11 21 47.1 &  +11 44 18.2 & 13.50 & 14.48 & 0.050 & AGN \nl 
 J112349.2+723002 & 11 23 51.8 &  +72 30 08.4 & 15.00 & 15.09 &	   &	 \nl   
 J112842.9+633559 & 11 28 41.6 &  +63 35 50.5 & 15.40 & &	   &	 \nl 
 J112850.7+231036 & 11 28 51.1 &  +23 10 37.0 & 15.40 & &	   &	 \nl   
 J112854.0+210630 & 11 28 55.2 &  +21 06 30.9 & 15.30 & &	   &	 \nl   
 J113109.0+263212 & 11 31 09.3 &  +26 32 07.8 & 15.40 & &	   &	 \nl   
 J113302.0+184655 & 11 33 02.0 &  +18 47 32.8 & 15.20 & &	   &	 \nl   
 J113313.3+500837 & 11 33 12.7 &  +50 08 56.6 & 15.00 & & 0.310 & GAL\nl 
 J113630.9+673708 & 11 36 30.1 &  +67 37 04.0 & 15.40 & & 0.135 & BL\nl  
 J113737.4+103931 & 11 37 38.1 &  +10 39 30.2 & 15.30 & &	   &	 \nl   
 J113826.8+032210 & 11 38 27.1 &	+3 22 09.9 & 14.90 & &	   &	 \nl   
 J113849.7+574245 & 11 38 49.6 &  +57 42 43.9 & 13.60 & & 0.115 & AGN \nl 
 J114009.0+030727 & 11 40 08.7 &	+3 07 11.0 & 15.30 & &	   &	 \nl 
 J114106.1+024110 & 11 41 05.7 &	+2 41 16.3 & 15.10 & &	   &	 \nl   
 J114247.5+215717 & 11 42 45.8 &  +21 57 22.4 & 15.30 & 13.25 &	   &	 \nl 
 J114509.2+381326 & 11 45 09.9 &  +38 13 29.1 & 15.40 & &	   &	 \nl 
 J114509.3+304724 & 11 45 10.3 &  +30 47 16.7 & 14.30 & & 0.059 & AGN \nl 
 J114606.1+035959 & 11 46 06.2 &	+3 59 55.2 & 14.60 & &	   &	 \nl   
 J114755.3+090235 & 11 47 55.0 &	+9 02 28.6 & 14.50 & &	   &	 \nl   
 J115137.3+561341 & 11 51 38.1 &  +56 13 30.8 & 15.10 & &	   &	 \nl   
 J115553.6+732416 & 11 55 54.2 &  +73 23 44.7 & 15.00 & 15.65 &	   &	 \nl   
 J115719.0+333645 & 11 57 17.4 &  +33 36 39.9 & 15.10 & & 0.213 & GAL\nl
 J115746.9+412642 & 11 57 46.1 &  +41 26 37.4 & 14.40 & &	   &	 \nl   
 J120333.4+022939 & 12 03 32.9 &	+2 29 34.5 & 14.40 & 13.88 & 0.077 & AGN \nl 
 J120547.8+584828 & 12 05 48.9 &  +58 48 30.0 & 15.20 & &	   &	 \nl   
 J120954.9+062806 & 12 09 54.6 &	+6 28 13.3 & 14.40 & &	   &	 \nl 
 J121104.0+700536 & 12 11 03.9 &  +70 05 31.3 & 14.00 & 14.59 &	   &	 \nl   
 J121157.1+055800 & 12 11 57.5 &	+5 58 00.9 & 14.10 & &	   &	 \nl   
 J121217.1+280356 & 12 12 16.2 &  +28 04 07.4 & 15.40 & & 0.167 & AGN \nl 
 J121417.7+140312 & 12 14 17.7 &  +14 03 12.6 & 13.90 & 13.96 & 0.081 & AGN \nl 
 J121510.9+073205 & 12 15 10.9 &	+7 32 03.8 & 15.40 & & 0.136 & BL\nl  
 J121752.1+300705 & 12 17 52.0 &  +30 06 59.9 & 14.40 & & 0.237 & BL\nl  
 J122044.5+690533 & 12 20 47.9 &  +69 05 37.7 & 15.00 & &	 * & GAL \nl
 J122144.4+751848 & 12 21 44.0 &  +75 18 38.5 & 14.80 & 14.37 & 0.070 & GAL\nl 
 J122523.1+042128 & 12 25 22.9 &	+4 21 18.6 & 14.10 & &  & \nl 
 J122623.7+372657 & 12 26 23.3 &  +37 27 01.3 & 15.40 & &	   &	 \nl   
 J122635.9+455933 & 12 26 36.9 &  +45 59 40.4 & 15.10 & &	   &	 \nl   
 J122745.1+084147 & 12 27 44.8 &	+8 41 49.8 & 13.60 & & 0.084 & AGN \nl 
 J122859.5+272527 & 12 29 00.3 &  +27 25 21.4 & 15.10 & &	   &	 \nl 
 J123132.5+641420 & 12 31 31.3 &  +64 14 17.5 & 15.00 & & 0.170 & BL\nl  
 J123154.6+323248 & 12 31 55.1 &  +32 32 40.8 & 14.40 & &	   &	 \nl   
 J123235.8+060315 & 12 32 35.8 &	+6 03 09.7 & 15.10 & &	   &	 \nl   
 J123325.8+093119 & 12 33 25.8 &	+9 31 23.0 & 14.10 & & 0.415 & AGN \nl 
 J123658.8+631111 & 12 36 58.7 &  +63 11 12.9 & 15.20 & &  & GAL\nl
 J123942.4+342453 & 12 39 42.5 &  +34 24 55.7 & 15.20 & &	   &	 \nl   
 J124129.4+372206 & 12 41 29.4 &  +37 22 01.7 & 13.90 & & 0.063 & AGN \nl 
 J124141.2+344032 & 12 41 39.9 &  +34 40 17.6 & 14.60 & 14.96 &	& \nl
 J124211.3+331703 & 12 42 10.6 &  +33 17 02.2 & 14.10 & 14.98 & 0.044 & AGN \nl 
 J124306.2+421233 & 12 43 07.1 &  +42 12 31.1 & 15.30 & &	   &	 \nl   
 J124306.5+353859 & 12 43 04.2 &  +35 39 16.8 & 14.90 & 15.23 &   & \nl 
 J124324.2+271645 & 12 43 24.7 &  +27 16 48.6 & 14.70 & &  & GAL\nl
 J124339.6+700517 & 12 43 39.3 &  +70 05 29.9 & 14.70 & 15.68 & & \nl   
 J124701.3+442325 & 12 47 00.1 &  +44 23 13.7 & 15.30 & &	   &	 \nl   
 J124717.2+481240 & 12 47 16.3 &  +48 12 39.6 & 15.30 & &	   &	 \nl 
 J124818.9+582031 & 12 48 18.7 &  +58 20 28.8 & 14.50 & &	   & BL\nl
 J125005.7+263118 & 12 50 05.7 &  +26 31 07.3 & 15.20 & & 2.043 & AGN \nl 
 J125422.6+793618 & 12 54 23.1 &  +79 36 12.8 & 15.20 & &	   &	 \nl   
 J125801.0+470237 & 12 57 59.4 &  +47 02 01.3 & 14.80 & &	   &	 \nl 
 J125830.1+652121 & 12 58 27.8 &  +65 21 30.7 & 15.40 & &  & \nl 
 J125851.4+235532 & 12 58 51.4 &  +23 55 26.6 & 13.90 & & 0.071 & AGN \nl 
 J130052.9+564101 & 13 00 50.7 &  +56 40 51.1 & 15.40 & &	   &	 \nl 
 J130258.8+162423 & 13 02 58.8 &  +16 24 27.7 & 14.90 & 15.09 & 0.067 & AGN \nl 
 J130425.4+333512 & 13 04 27.2 &  +33 35 13.0 & 14.40 & & 0.188 & GAL\nl
 J130803.0+035124 & 13 08 03.1 &	+3 51 14.1 & 15.10 & &	   &	 \nl   
 J130947.1+081949 & 13 09 47.0 &	+8 19 48.9 & 14.80 & & 0.155 & AGN \nl 
 J131218.0+351524 & 13 12 17.7 &  +35 15 20.3 & 14.70 & & 0.184 & AGN \nl 
 J131334.0+725914 & 13 13 32.0 &  +72 59 10.9 & 15.20 & & 0.112 & AGN \nl 
 J131349.6+365357 & 13 13 49.0 &  +36 53 57.7 & 15.40 & &	   &	 \nl   
 J131414.6+412347 & 13 14 18.3 &  +41 24 30.1 & 15.30 & &	   &	 \nl   
 J131432.5+122706 & 13 14 32.7 &  +12 27 17.9 & 14.70 & &	   &	 \nl   
 J131451.5+421819 & 13 14 51.5 &  +42 18 19.1 & 14.80 & &	   &	 \nl 
 J131555.1+212508 & 13 15 55.1 &  +21 25 21.5 & 15.00 & &	   &	 \nl 
 J131750.4+601047 & 13 17 50.3 &  +60 10 40.6 & 15.30 & &	   &	 \nl   
 J132025.1+690018 & 13 20 24.6 &  +69 00 12.4 & 15.40 & & 0.067 & AGN \nl
 J132042.4+601526 & 13 20 45.3 &  +60 15 16.2 & 14.00 & &       &     \nl 
 J132314.2+463132 & 13 23 14.9 &  +46 31 21.8 & 15.00 & & 0.143 & AGN \nl 
 J132400.2+573918 & 13 24 00.8 &  +57 39 16.1 & 13.90 & & 0.115 & BL\nl
 J132434.9+475802 & 13 24 35.5 &  +47 58 00.7 & 15.10 & &	   &	 \nl   
 J132602.2+601206 & 13 26 02.3 &  +60 11 59.4 & 15.10 & &	   &	 \nl 
 J132632.2+792850 & 13 26 32.3 &  +79 28 51.7 & 15.00 & & & \nl 
 J132847.3+503808 & 13 28 48.5 &  +50 37 53.5 & 15.30 & &	   &	 \nl   
 J132908.3+295018 & 13 29 08.8 &  +29 50 23.9 & 15.40 & & 0.047 & AGN \nl 
 J132943.8+315338 & 13 29 43.6 &  +31 53 36.3 & 14.60 & & 0.090 & AGN \nl 
 J133434.3+575019 & 13 34 35.3 &  +57 50 15.3 & 15.00 & &	   &	 \nl 
 J133439.6+171748 & 13 34 37.3 &  +17 17 49.4 & 15.30 & &	   &	 \nl   
 J133608.2+755041 & 13 36 09.8 &  +75 50 34.9 & 15.20 & 15.28 &	   &	 \nl   
 J133718.8+242306 & 13 37 18.7 &  +24 23 02.9 & 14.30 & 14.26 & 0.107 & AGN \nl 
 J133826.6+321252 & 13 38 26.9 &  +32 12 51.9 & 15.20 & &	   &	 \nl   
 J133908.5+115855 & 13 39 08.5 &  +11 58 53.5 & 15.30 & &	   &	 \nl   
 J133938.5+183055 & 13 39 37.8 &  +18 30 59.4 & 15.10 & &	   &	 \nl   
 J134021.4+274100 & 13 40 21.9 &  +27 41 26.8 & 14.20 & 14.19 & & \nl 
 J134210.9+564219 & 13 42 10.1 &  +56 42 10.9 & 14.60 & & 0.040 & AGN \nl 
 J134335.3+413839 & 13 43 35.7 &  +41 38 24.3 & 14.70 & 14.60 &	   &	 \nl 
 J134356.7+253845 & 13 43 56.7 &  +25 38 46.9 & 14.10 & 15.09 &  & \nl 
 J134357.3+271252 & 13 43 57.4 &  +27 12 40.9 & 15.40 & & 0.077 & AGN \nl 
 J134453.1+000525 & 13 44 52.9 &	+0 05 19.7 & 14.80 & 15.51 &	   &	 \nl   
 J134607.5+293814 & 13 46 08.1 &  +29 38 10.5 & 14.10 & & 0.076 & AGN \nl 
 J135022.2+094007 & 13 50 22.1 &	+9 40 10.7 & 14.00 & &	   &	 \nl   
 J135022.2+094007 & 13 50 22.1 &	+9 40 10.7 & 14.00 & &	   &	 \nl   
 J135143.8+242420 & 13 51 43.9 &  +24 24 21.5 & 14.90 & &	   &	 \nl   
 J135436.0+180523 & 13 54 35.6 &  +18 05 17.2 & 15.30 & & 0.152 & AGN \nl 
 J135553.3+383427 & 13 55 53.5 &  +38 34 29.1 & 15.00 & & 0.051 & AGN \nl 
 J135821.2+360356 & 13 58 24.5 &  +36 03 47.7 & 15.00 & 15.36 &	   &	 \nl   
 J140310.5+375810 & 14 03 08.8 &  +37 58 27.6 & 15.20 & &	   &	 \nl   
 J140519.6+020008 & 14 05 19.4 &	+2 00 05.2 & 14.80 & & 0.000 & STAR\nl  
 J140606.1+580045 & 14 06 04.8 &  +58 00 41.3 & 15.30 & &	   &	 \nl   
 J140622.2+222350 & 14 06 21.9 &  +22 23 46.7 & 15.10 & & 0.098 & AGN \nl 
 J140924.1+261827 & 14 09 23.9 &  +26 18 21.3 & 15.40 & & 0.940 & AGN \nl 
 J141336.8+702954 & 14 13 36.7 &  +70 29 50.4 & 14.30 & & 0.107 & AGN \nl 
 J141342.6+433938 & 14 13 43.7 &  +43 39 44.1 & 14.70 & & 0.089 & GAL \nl
 J141346.6+263246 & 14 13 45.3 &  +26 33 03.1 & 15.00 & 14.85 &	   &	 \nl   
 J141700.5+445556 & 14 17 00.8 &  +44 56 06.0 & 14.70 & 15.17 & 0.114 & AGN \nl 
 J141756.8+254329 & 14 17 56.7 &  +25 43 24.7 & 15.30 & & 0.237 & BL \nl
 J141758.8+360749 & 14 17 58.0 &  +36 08 10.5 & 15.20 & &	   &	 \nl   
 J141901.9+280942 & 14 19 01.9 &  +28 09 41.7 & 14.80 & &	   &	 \nl   
 J142058.6+262450 & 14 20 56.1 &  +26 24 22.5 & 15.40 & 14.95 &	   &	 \nl   
 J142107.1+253818 & 14 21 07.6 &  +25 38 20.8 & 15.40 & & 1.050 & AGN \nl 
 J142129.8+474719 & 14 21 29.8 &  +47 47 24.7 & 14.30 & 14.62 & 0.072 & AGN \nl 
 J142313.4+505537 & 14 23 14.3 &  +50 55 38.1 & 15.20 & & 0.274 & AGN \nl 
 J142425.2+595254 & 14 24 24.1 &  +59 53 00.7 & 15.10 & 14.18 &	   &	 \nl   
 J142630.6+390348 & 14 26 30.7 &  +39 03 43.5 & 13.50 & &	   &	 \nl   
 J142700.5+234803 & 14 27 00.4 &  +23 48 00.1 & 14.80 & &	   &   BL\nl
 J142725.3+194954 & 14 27 25.0 &  +19 49 52.3 & 14.00 & 15.60 & 0.131 & AGN \nl 
 J142906.7+011708 & 14 29 06.5 &	+1 17 05.0 & 13.60 & 13.21 & 0.086 & AGN \nl 
 J142924.3+451826 & 14 29 25.0 &  +45 18 31.6 & 15.10 & &	   &	 \nl   
 J143308.8+232650 & 14 33 08.4 &  +23 26 31.1 & 15.10 & &	   &	 \nl 
 J143445.8+332814 & 14 34 45.3 &  +33 28 19.8 & 14.70 & &	   &	 \nl 
 J144034.4+242255 & 14 40 34.3 &  +24 22 50.4 & 15.30 & &	   &	 \nl 
 J144248.5+120042 & 14 42 48.2 &  +12 00 40.4 & 14.60 & & 0.162 & BL\nl
 J144645.8+403510 & 14 46 45.9 &  +40 35 05.8 & 15.10 & & 0.267 & AGN \nl 
 J144754.0+283323 & 14 47 54.2 &  +28 33 23.7 & 15.40 & &	   &	 \nl   
 J144825.6+355955 & 14 48 25.0 &  +35 59 46.4 & 15.00 & & 0.111 & AGN \nl 
 J145307.6+255438 & 14 53 08.0 &  +25 54 32.8 & 15.40 & &	   &	 \nl   
 J145307.8+215333 & 14 53 08.3 &  +21 53 38.5 & 15.10 & &	   &	 \nl   
 J145559.0+492158 & 14 55 59.5 &  +49 21 52.3 & 15.40 & &	   &	 \nl   
 J145729.4+083356 & 14 57 29.0 &	+8 34 22.6 & 15.00 & & 0.167 & AGN \nl 
 J145843.1+213614 & 14 58 42.7 &  +21 36 10.0 & 15.30 & & 0.062 & AGN \nl 
 J150023.0+763644 & 15 00 22.3 &  +76 36 37.7 & 14.70 & 15.14 &	   &	 \nl   
 J150124.1+302638 & 15 01 24.2 &  +30 26 33.2 & 15.30 & &	   &	 \nl 
 J150317.5+681011 & 15 03 16.3 &  +68 10 05.7 & 15.00 & &       &     \nl 
 J150332.0+295026 & 15 03 32.1 &  +29 50 23.9 & 14.80 & &	   &	 \nl   
 J150506.8+435002 & 15 05 07.3 &  +43 50 05.1 & 15.00 & &	   &	 \nl 
 J150752.3+515116 & 15 07 52.6 &  +51 51 11.1 & 15.40 & &	   &	 \nl   
 J151040.8+333515 & 15 10 41.1 &  +33 35 05.4 & 15.30 & &	   &	 \nl 
 J151105.3+525128 & 15 11 06.0 &  +52 51 26.8 & 15.00 & &	   &	 \nl 
 J151447.0+351348 & 15 14 46.9 &  +35 13 48.6 & 14.80 & &	   &	 \nl   
 J151634.5+205847 & 15 16 34.5 &  +20 58 37.4 & 14.90 & &	   &	 \nl   
 J151845.3+061340 & 15 18 45.7 &	+6 13 55.8 & 14.40 & & 0.102 & AGN \nl 
 J151921.7+590823 & 15 19 21.6 &  +59 08 23.6 & 14.40 & 15.09 & 0.078 & AGN \nl 
 J152558.6+181423 & 15 25 58.5 &  +18 14 15.6 & 15.00 & &	   &	 \nl   
 J152806.5+132337 & 15 28 06.9 &  +13 23 50.3 & 15.20 & &	   &	 \nl   
 J152912.9+381226 & 15 29 14.0 &  +38 13 06.0 & 15.00 & 15.06 & 0.000 & STAR\nl  
 J153140.9+201927 & 15 31 41.3 &  +20 19 30.1 & 15.30 & &	   &	 \nl   
 J153202.3+301631 & 15 32 02.2 &  +30 16 28.6 & 13.50 & 15.30 & 0.064 & BL\nl
 J153718.8+084355 & 15 37 20.5 &	+8 44 08.7 & 15.20 & &	   &	 \nl 
 J153935.2+473545 & 15 39 34.9 &  +47 35 53.1 & 15.00 & 14.87 &       &     \nl 
 J154236.8+581153 & 15 42 36.9 &  +58 11 45.0 & 13.90 & 14.07 &	   &	 \nl   
 J154508.1+170935 & 15 45 07.5 &  +17 09 50.4 & 14.70 & & 0.045 & AGN \nl 
 J154732.3+102446 & 15 47 32.2 &  +10 24 51.2 & 15.40 & &	   &	 \nl   
 J154751.7+025538 & 15 47 51.9 &	+2 55 50.8 & 14.70 & & 0.098 & AGN \nl 
 J154814.6+450040 & 15 48 14.7 &  +45 00 27.8 & 14.90 & &	   &	 \nl   
 J155023.8+281125 & 15 50 24.0 &  +28 11 17.2 & 15.20 & &	   &	 \nl   
 J155041.6+413915 & 15 50 39.0 &  +41 39 29.9 & 15.30 & &	   &	 \nl 
 J155411.8+241415 & 15 54 10.9 &  +24 14 40.5 & 15.40 & &	   &	 \nl   
 J155444.6+082202 & 15 54 44.6 &	+8 22 21.6 & 15.40 & & 0.119 & AGN \nl 
 J155543.2+111114 & 15 55 43.0 &  +11 11 24.1 & 14.30 & 13.81 & 0.360 & BL\nl
 J155643.0+294838 & 15 56 42.8 &  +29 48 47.5 & 13.90 & 14.63 & 0.087 & AGN \nl 
 J155745.0+353020 & 15 57 42.3 &  +35 30 29.9 & 13.90 & &	   &	 \nl   
 J155818.7+255118 & 15 58 18.8 &  +25 51 24.4 & 15.30 & & 0.070 & AGN \nl 
 J160529.2+720852 & 16 05 26.0 &  +72 08 36.3 & 15.20 & &	   &	 \nl   
 J160740.7+254106 & 16 07 40.2 &  +25 41 12.6 & 13.80 & 13.87 &	   &	 \nl   
 J161047.7+330329 & 16 10 47.8 &  +33 03 37.7 & 14.10 & & 0.097 & AGN \nl 
 J161413.0+260412 & 16 14 13.2 &  +26 04 15.9 & 15.10 & & 0.131 & AGN \nl 
 J161601.3+323222 & 16 16 01.8 &  +32 32 28.8 & 15.10 & & 0.118 & AGN \nl 
 J161711.4+063816 & 16 17 10.5 &	+6 38 43.0 & 15.20 & & 0.092 & AGN \nl 
 J161804.5+672409 & 16 18 03.8 &  +67 23 50.0 & 15.00 & &	   &	 \nl 
 J161809.2+361951 & 16 18 09.4 &  +36 19 57.8 & 13.80 & & 0.034 & AGN \nl 
 J161814.2+293828 & 16 18 14.0 &  +29 38 08.9 & 15.30 & &	   &	 \nl   
 J162011.5+172413 & 16 20 11.3 &  +17 24 27.5 & 15.20 & & 0.114 & AGN \nl 
 J162100.4+254547 & 16 21 00.3 &  +25 46 03.3 & 14.70 & &	   &	 \nl   
 J162114.3+181936 & 16 21 14.4 &  +18 19 49.9 & 15.20 & & 0.125 & AGN \nl 
 J162348.2+402948 & 16 23 48.2 &  +40 29 59.0 & 15.00 & &	   &	 \nl   
 J162355.9+370018 & 16 23 56.4 &  +37 00 44.9 & 15.30 & &	   &	 \nl   
 J162456.7+755457 & 16 24 56.5 &  +75 54 55.8 & 13.70 & & 0.200 & AGN \nl 
 J162607.6+335902 & 16 26 07.2 &  +33 59 15.0 & 14.80 & & 0.204 & AGN \nl 
 J163116.3+095545 & 16 31 16.0 &	+9 55 57.9 & 15.00 & & 0.092 & AGN \nl 
 J163323.3+471848 & 16 33 23.5 &  +47 19 00.1 & 14.60 & & 0.116 & AGN \nl 
 J163338.4+371311 & 16 33 38.7 &  +37 13 14.8 & 14.70 & &  & \nl 
 J163509.5+343956 & 16 35 09.2 &  +34 40 03.4 & 15.40 & &	   &	 \nl   
 J163523.2+545304 & 16 35 23.2 &  +54 53 00.3 & 15.00 & &	   &	 \nl   
 J164443.2+261909 & 16 44 44.1 &  +26 19 04.6 & 15.20 & &       &     \nl 
 J164550.2+792129 & 16 45 49.5 &  +79 21 28.6 & 14.90 & &	   &	 \nl   
 J164625.8+392922 & 16 46 26.0 &  +39 29 32.2 & 14.60 & & 0.100 & AGN \nl 
 J164735.4+495001 & 16 47 34.8 &  +49 49 59.8 & 14.60 & & 0.047 & AGN \nl 
 J164801.1+295650 & 16 48 00.8 &  +29 56 57.4 & 14.30 & & 0.101 & AGN \nl 
 J165141.2+721824 & 16 51 39.6 &  +72 18 42.7 & 15.40 & &	   &	 \nl   
 J165253.7+400927 & 16 52 56.6 &  +40 08 43.1 & 15.00 & &	   &	 \nl 
 J170328.3+614114 & 17 03 28.9 &  +61 41 10.1 & 14.70 & &	   &	 \nl   
 J170425.2+333145 & 17 04 22.4 &  +33 31 40.3 & 14.90 & 13.46 &	   &	 \nl   
 J170535.1+334011 & 17 05 34.9 &  +33 40 12.3 & 14.90 & &	   &	 \nl   
 J171013.2+334410 & 17 10 13.5 &  +33 44 03.6 & 14.80 & & 0.208 & AGN \nl 
 J171322.8+325631 & 17 13 22.6 &  +32 56 28.8 & 14.50 & & 0.100 & AGN \nl 
 J171410.8+575826 & 17 14 11.5 &  +57 58 33.5 & 14.90 & & 0.092 & AGN \nl 
 J171601.3+311215 & 17 16 01.9 &  +31 12 13.5 & 14.40 & 15.48 & 0.111 & AGN \nl 
 J171935.9+424518 & 17 19 33.9 &  +42 45 22.5 & 15.00 & &	   &	 \nl   
 J172320.5+341756 & 17 23 20.8 &  +34 17 57.8 & 14.50 & & 0.206 & AGN\nl 
 J172609.3+743103 & 17 26 08.3 &  +74 31 03.4 & 14.60 & & 0.052 & AGN \nl 
 J172855.8+515654 & 17 28 54.6 &  +51 56 49.2 & 14.90 & 14.21 &  & \nl 
 J173114.5+323250 & 17 31 15.2 &  +32 32 58.4 & 13.90 & 14.05 & & \nl   
 J174025.8+514942 & 17 40 25.7 &  +51 49 42.4 & 14.50 & &  &  \nl 
 J174815.0+582333 & 17 48 15.3 &  +58 23 35.5 & 15.00 & &	   &	 \nl   
 J174839.6+530240 & 17 48 37.6 &  +53 02 45.4 & 15.40 & &	   &	 \nl   
 J214923.8+092921 & 21 49 23.7 &	+9 28 47.3 & 15.30 & &	   &	 \nl 
 J215912.7+095247 & 21 59 12.3 &	+9 52 43.4 & 14.90 & & 0.101 & AGN \nl 
 J222602.8+172245 & 22 26 02.2 &  +17 22 47.0 & 15.40 & &  & \nl 
 J224939.6+110016 & 22 49 39.6 &  +11 00 29.2 & 14.80 & & 0.084 & AGN \nl 
 J225207.7+145448 & 22 52 08.1 &  +14 54 49.6 & 14.80 & & 0.130 & AGN \nl 
 J225636.8+052522 & 22 56 36.5 &	+5 25 17.2 & 14.90 & & 0.066 & AGN \nl 
 J225932.9+245505 & 22 59 32.9 &  +24 55 05.6 & 13.50 & 15.07 & 0.034 & AGN \nl 
 J231357.3+144424 & 23 13 56.3 &  +14 43 53.5 & 14.90 & &	   &	 \nl   
 J231517.5+182825 & 23 15 17.1 &  +18 28 14.4 & 15.00 & & 0.104 & AGN \nl 
 J232339.1+090842 & 23 23 39.0 &   +9 08 50.6 & 14.80 & & 0.068 & AGN \nl 
 J233606.6+241555 & 23 36 06.1 &  +24 15 58.3 & 14.50 & & 0.039 & AGN \nl 
 J233641.8+235526 & 23 36 42.2 &  +23 55 29.0 & 14.20 & & 0.127 & GAL\nl
 J233739.8+001604 & 23 37 40.7 &	+0 16 35.3 & 14.90 & &       &     \nl
 J234031.5+102934 & 23 40 31.0 &  +10 29 39.0 & 14.70 & &	   &	 \nl   
 J234339.0+024445 & 23 43 39.8 &	+2 45 03.9 & 15.40 & & 0.091 & AGN \nl 
 J235257.1+032008 & 23 52 58.0 &	+3 20 17.3 & 14.30 & & 0.086 & AGN \nl 
 J235754.3+132418 & 23 57 53.8 &  +13 24 09.6 & 15.30 & &	   &	 \nl 
\enddata
\label{tbl-2a}
\end{deluxetable}

\begin{deluxetable}{lrrrrc}
\tablenum{2}
\clearpage
\tablewidth{0pt}
\tablecaption{THE GSC SAMPLE}
\tablehead{
\colhead{Name (1RXS)}           & \colhead{R.A.} 	&
\colhead{Declination}           & \colhead{$V_{GSC}$}       &
\colhead{$z$}                   & \colhead{Type}}       
\startdata
 J000350.4+020340 & 00 03 49.7 &	+2 03 58.9 &  13.28 &  & \nl
 J001219.0+100602 & 00 12 19.3 &  +10 06 45.8 &  14.02 & 	   & 	   \nl
 J003633.7+254513 & 00 36 32.4 &  +25 45 18.2 &  14.32 & 	   & 	   \nl
 J004400.0+313729 & 00 44 00.0 &  +31 37 04.3 &  14.42 & 	   & 	   \nl
 J004719.4+144215 & 00 47 19.4 &  +14 42 11.8 &  14.00 &  0.039 &  AGN \nl
 J004931.6+112832 & 00 49 32.0 &  +11 28 26.0 &  14.37 &  0.275 &  AGN \nl
 J005017.9+083734 & 00 50 17.4 &	+8 37 35.3 &  14.31 & 	   & 	   \nl
 J005029.2+112902 & 00 50 27.9 &  +11 29 10.9 &  14.37 &  0.000 &  STAR\nl
 J005351.3+221222 & 00 53 50.9 &  +22 12 13.7 &  14.38 & 	   & 	   \nl
 J005953.3+314934 & 00 59 53.3 &  +31 49 37.4 &  13.74 &  0.015 &  AGN \nl
 J010014.0+055200 & 01 00 14.1 &	+5 51 54.8 &  14.00 & 	   & 	   \nl
 J011125.4+152625 & 01 11 24.8 &  +15 26 26.8 &  13.59 & 	   & 	   \nl
 J011704.2+000025 & 01 17 03.6 &	+0 00 27.0 &  14.50 &  0.040 &  AGN \nl
 J012732.9+191043 & 01 27 32.5 &  +19 10 43.8 &  11.80 &  0.017 &  AGN \nl
 J015240.2+014718 & 01 52 39.6 &	+1 47 16.8 &  13.92 &  0.080 & BL\nl
 J015242.1+010040 & 01 52 41.9 &   +1 00 25.1 &  13.57 & 0.230 & GAL\nl
 J020026.7+024012 & 02 00 26.3 &	+2 40 09.9 &  12.77 &  0.078 &  AGN\nl
 J024920.8+191813 & 02 49 20.7 &  +19 18 14.2 &  14.18 &  0.031 &  AGN\nl
 J025153.2+222735 & 02 51 53.7 &  +22 27 35.7 &  12.40 &  0.000 &  STAR\nl
 J075704.9+583245 & 07 57 06.2 &  +58 32 58.7 &  14.48 &  0.168 &  AGN\nl
 J080525.8+753424 & 08 05 23.7 &  +75 34 34.7 &  14.40 &        &     \nl
 J081228.3+623627 & 08 12 28.2 &  +62 36 23.3 &  14.15 & 	   & 	   \nl
 J081517.8+460429 & 08 15 16.9 &  +46 04 30.7 &  14.24 & 	   & 	   \nl
 J081917.9+642943 & 08 19 17.6 &  +64 29 40.0 &  14.40 &  0.039 &  AGN \nl
 J082407.3+613612 & 08 24 11.3 &  +61 36 11.3 &  14.01 &   & \nl
 J083137.6+192339 & 08 31 38.3 &  +19 23 45.2 &  11.43 & 	   & 	   \nl
 J083811.0+245336 & 08 38 10.9 &  +24 53 42.4 &  12.80 & 	   & 	   \nl
 J084456.2+425826 & 08 44 56.6 &  +42 58 35.1 &  14.40 & 	   & 	   \nl
 J084602.9+830757 & 08 46 17.9 &  +83 07 43.5 &  14.47 & 	   & 	   \nl
 J084742.5+344506 & 08 47 42.5 &  +34 45 03.8 &  13.64 &  0.064 &  AGN \nl
 J090008.1+743419 & 09 00 03.7 &  +74 34 26.4 &  14.37 & 	   & 	   \nl
 J091230.8+155531 & 09 12 31.0 &  +15 55 24.3 &  13.42 & 	   & 	   \nl
 J091552.3+090056 & 09 15 51.8 &	+9 00 50.9 &  12.99 & 0.000 & STAR \nl 
 J091826.2+161825 & 09 18 26.0 &  +16 18 19.2 &  13.21 &  0.030 &  AGN \nl
 J092030.8+013544 & 09 20 31.1 &	+1 35 37.1 &  14.14 & 	   & 	   \nl
 J092108.2+480201 & 09 21 11.3 &  +48 01 59.2 &  13.84 & 	   & 	   \nl
 J092343.0+225437 & 09 23 43.0 &  +22 54 32.7 &  13.43 &   &  \nl
 J092512.3+521716 & 09 25 13.0 &  +52 17 11.4 &  13.66 &  0.036 &  AGN \nl
 J092603.6+124406 & 09 26 03.3 &  +12 44 03.3 &  13.71 &  0.028 &  AGN \nl
 J092702.8+390221 & 09 27 04.0 &  +39 02 17.8 &  13.59 &      &     \nl
 J092705.7+374157 & 09 27 03.0 &  +37 42 05.5 &  14.12 & 	   &     \nl
 J093701.0+010548 & 09 37 01.0 &	+1 05 43.0 &  13.87 &  0.051 &  AGN \nl
 J093900.4+253008 & 09 39 00.4 &  +25 30 14.6 &  14.41 & 	   & 	   \nl
 J094204.0+234106 & 09 42 04.8 &  +23 41 06.5 &  14.15 & 0.021 & GAL\nl
 J094432.8+573544 & 09 44 04.7 &  +57 33 28.1 &  14.13 & 	   & 	   \nl
 J094851.0+153901 & 09 48 50.2 &  +15 38 34.6 &  14.25 & 	   & 	   \nl
 J095340.4+014154 & 09 53 41.3 &	+1 42 01.7 &  13.98 & 	   & 	   \nl
 J095624.5+064803 & 09 56 23.8 &	+6 48 01.6 &  14.34 & 	   & 	   \nl
 J095919.3+435033 & 09 59 19.2 &  +43 50 35.4 &  13.57 & 	   & 	   \nl
 J100446.0+144651 & 10 04 47.6 &  +14 46 45.2 &  14.23 &  0.082 &  AGN \nl
 J100641.5+213955 & 10 06 43.6 &  +21 39 27.7 &  14.50 & 	   & 	   \nl
 J101218.6+631133 & 10 12 21.6 &  +63 11 32.1 &  14.21 & 	   & 	   \nl
 J101718.0+291439 & 10 17 18.3 &  +29 14 33.8 &  13.96 & 0.048 & AGN\nl
 J101912.1+635802 & 10 19 12.6 &  +63 58 02.2 &  13.69 & 0.041 & AGN\nl
 J102334.6+443346 & 10 23 35.0 &  +44 33 41.5 &  13.86 & 	   & 	   \nl
 J102407.0+273130 & 10 24 06.7 &  +27 31 22.7 &  14.22 & 	   & 	   \nl
 J102611.1+523755 & 10 26 06.2 &  +52 37 56.3 &  13.99 &   & \nl
 J104038.7+373233 & 10 40 39.0 &  +37 32 31.3 &  13.36 & 	   & 	   \nl
 J104333.4+010109 & 10 43 32.8 &	+1 01 08.6 &  13.99 &  0.072 &  AGN \nl
 J104439.4+384541 & 10 44 39.1 &  +38 45 34.8 &  14.48 & 0.036 & GAL\nl
 J105121.3+360728 & 10 51 21.3 &  +36 07 27.4 &  13.31 & 	   & 	   \nl
 J105214.2+055514 & 10 52 15.3 &	+5 55 08.2 &  13.96 & 	   & 	   \nl
 J105328.5+053052 & 10 53 29.7 &	+5 30 30.2 &  13.92 & 	   & 	   \nl
 J105340.7+525310 & 10 53 41.2 &  +52 53 01.7 &  14.25 & 	   & 	   \nl
 J110159.1+572316 & 11 02 00.1 &  +57 22 50.3 &  14.43 & 	   & 	   \nl
 J110310.2+363911 & 11 03 10.7 &  +36 39 06.3 &  13.41 & 	   & 	   \nl
 J110321.2+133759 & 11 03 21.8 &  +13 37 52.4 &  12.90 & 	   & 	   \nl
 J110455.7+433421 & 11 04 56.0 &  +43 34 03.0 &  14.29 & 	   & 	   \nl
 J110943.6+214519 & 11 09 41.1 &  +21 44 24.2 &  14.33 & 0.032 & GAL\nl
 J111300.1+102518 & 11 13 00.2 &  +10 25 12.3 &  14.29 & 	   & 	   \nl
 J111349.5+093518 & 11 13 49.7 &	+9 35 10.9 &  12.42 &  0.029 &  AGN \nl
 J112147.3+114420 & 11 21 47.1 &  +11 44 18.2 &  14.48 &  0.050 &  AGN \nl
 J112150.8+405147 & 11 21 51.2 &  +40 51 46.4 &  13.97 & 	   & 	   \nl
 J112315.6+193610 & 11 23 14.6 &  +19 35 25.5 &  14.13 & 	   & 	   \nl
 J112536.7+542243 & 11 25 36.1 &  +54 22 56.9 &  14.33 &  0.021 &  AGN\nl
 J114116.2+215624 & 11 41 16.2 &  +21 56 21.1 &  13.25 &  0.063 &  AGN\nl
 J114516.1+794054 & 11 45 16.1 &  +79 40 52.6 &  13.50 &  0.065 &  AGN\nl
 J114738.0+050119 & 11 47 37.4 &	+5 01 09.3 &  12.09 & 	   & 	   \nl
 J114741.4+001524 & 11 47 41.7 &	+0 15 24.1 &  14.23 &  & \nl
 J115658.6+241523 & 11 56 55.8 &  +24 15 35.2 &  13.55 & 0.142 & GAL\nl
 J120333.4+022939 & 12 03 32.9 &	+2 29 34.5 &  13.88 &  0.077 &  AGN\nl
 J120829.9+132752 & 12 08 29.8 &  +13 28 06.0 &  13.31 & 	   & 	   \nl
 J121417.7+140312 & 12 14 17.7 &  +14 03 12.6 &  13.96 &  0.081 &  AGN\nl
 J121607.4+504926 & 12 16 07.0 &  +50 49 30.2 &  14.25 &  0.031 &  AGN\nl
 J121900.7+110727 & 12 18 59.8 &  +11 07 53.9 &  13.53 & 	   & 	   \nl
 J121920.9+063838 & 12 19 21.5 &	+6 38 43.9 &  13.22 &     &    \nl
 J122005.9+650552 & 12 20 10.3 &  +65 05 55.2 &  14.34 & 	   & 	   \nl
 J122144.4+751848 & 12 21 44.0 &  +75 18 38.5 &  14.37 & 0.070 & AGN\nl
 J122147.1+015637 & 12 21 46.6 &	+1 56 35.3 &  13.76 & 	   & 	   \nl
 J122306.6+103722 & 12 23 06.7 &  +10 37 16.8 &  12.25 & 0.026 & GAL\nl
 J122324.4+024040 & 12 23 24.2 &	+2 40 44.9 &  12.81 &  0.023 &  AGN\nl
 J122512.5+321354 & 12 25 13.1 &  +32 14 00.9 &  14.38 &  0.061 &  AGN\nl
 J122906.5+020311 & 12 29 06.7 &	+2 03 08.1 &  12.26 &  0.158 &  AGN\nl
 J123014.2+251805 & 12 30 14.2 &  +25 18 05.9 &  14.49 &  0.135 & BL\nl
 J123055.5+315207 & 12 30 55.8 &  +31 52 16.1 &  14.19 & 	   & 	   \nl
 J123203.6+200930 & 12 32 03.6 &  +20 09 29.6 &  12.87 &  0.064 &  AGN\nl
 J123415.2+481306 & 12 34 16.0 &  +48 13 06.9 &  14.31 & 	   & 	   \nl
 J123651.1+453907 & 12 36 51.2 &  +45 39 04.4 &  13.59 &  0.290 &  AGN\nl
 J123658.6+455341 & 12 36 57.0 &  +45 53 26.0 &  14.41 &   & \nl
 J124147.5+564506 & 12 41 46.7 &  +56 45 13.5 &  13.13 & 	   & 	   \nl
 J124312.5+362743 & 12 43 12.7 &  +36 27 43.8 &  11.42 & 	   & 	   \nl
 J124955.0+102312 & 12 49 54.5 &  +10 23 08.3 &  14.03 & 	   & 	   \nl
 J125731.7+354313 & 12 57 32.7 &  +35 43 19.9 &  14.19 &      &      \nl
 J130934.9+285908 & 13 09 36.1 &  +28 59 15.0 &  13.95 & 	   & 	   \nl
 J131957.2+523533 & 13 19 58.8 &  +52 35 27.7 &  13.49 & 	   & 	   \nl
 J132016.3+330828 & 13 20 14.7 &  +33 08 36.1 &  13.17 & 0.036 & GAL\nl
 J133451.1+374616 & 13 34 51.9 &  +37 46 20.7 &  13.83 & 	   & 	   \nl
 J133718.8+242306 & 13 37 18.7 &  +24 23 02.9 &  14.26 &  0.107 &  AGN\nl
 J133752.7+204634 & 13 37 50.9 &  +20 46 39.8 &  14.20 & 	   & 	   \nl
 J134021.4+274100 & 13 40 21.9 &  +27 41 26.8 &  14.19 &   & \nl
 J134952.7+020446 & 13 49 52.8 &	+2 04 44.3 &  13.32 &   &  \nl
 J135119.8+033722 & 13 51 20.2 &	+3 37 16.4 &  13.77 & 	   & 	   \nl
 J135304.8+691832 & 13 53 03.4 &  +69 18 29.5 &  12.92 & 	   & 	   \nl
 J135420.2+325547 & 13 54 20.0 &  +32 55 47.9 &  12.78 &  0.026 &  AGN\nl
 J140226.8+054103 & 14 02 26.3 &	+5 40 51.8 &  13.26 & 	   & 	   \nl
 J141722.1+452544 & 14 17 21.9 &  +45 25 46.7 &  13.94 & 	   & 	   \nl
 J141759.6+250817 & 14 17 59.2 &  +25 08 13.2 &  11.02 &   &  \nl
 J141802.6+800710 & 14 17 59.3 &  +80 07 02.2 &  14.19 & 	   & 	   \nl
 J142425.2+595254 & 14 24 24.1 &  +59 53 00.7 &  14.18 & 	   & 	   \nl
 J142906.7+011708 & 14 29 06.5 &	+1 17 05.0 &  13.21 &  0.086 &  AGN\nl
 J143104.8+281716 & 14 31 04.8 &  +28 17 14.8 &  13.43 &  0.046 &  AGN\nl
 J143452.3+483938 & 14 34 52.4 &  +48 39 42.6 &  13.38 & 	   & 	   \nl
 J143729.6+412842 & 14 37 29.9 &  +41 28 35.2 &  13.70 & 	   & 	   \nl
 J144713.2+570205 & 14 47 13.0 &  +57 01 56.7 &  13.97 & 	   & 	   \nl
 J150401.5+102620 & 15 04 01.2 &  +10 26 16.2 &  14.19 &  0.036 &  AGN\nl
 J150406.7+485856 & 15 04 07.1 &  +48 58 55.1 &  14.05 & 	   & 	   \nl
 J150724.6+433356 & 15 07 23.5 &  +43 33 51.6 &  13.64 & 	   & 	   \nl
 J150950.2+415540 & 15 09 49.7 &  +41 55 38.8 &  14.29 & 	   & 	   \nl
 J151750.8+050615 & 15 17 51.7 &   +5 06 27.4 &  13.98 &  0.039 &  AGN\nl
 J151837.9+404506 & 15 18 38.9 &  +40 45 00.0 &  14.19 &   &  \nl
 J153345.9+690037 & 15 33 44.7 &  +69 00 34.0 &  13.73 & 	   & 	   \nl
 J153412.6+625902 & 15 34 13.2 &  +62 58 57.7 &  13.76 & 	   & 	   \nl
 J153522.9+600515 & 15 35 24.5 &  +60 05 15.3 &  13.17 & 	   & 	   \nl
 J153552.0+575404 & 15 35 52.4 &  +57 54 08.5 &  13.91 &  0.030 &  AGN\nl
 J153704.2+374830 & 15 37 04.0 &  +37 48 26.9 &  13.43 & 	   & 	   \nl
 J153944.2+275113 & 15 39 43.9 &  +27 50 58.1 &  13.74 & 	   & 	   \nl
 J154236.8+581153 & 15 42 36.9 &  +58 11 45.0 &  14.07 & 	   & 	   \nl
 J154348.9+401343 & 15 43 50.5 &  +40 13 41.6 &  13.05 & 	   & 	   \nl
 J154532.3+420500 & 15 45 34.7 &  +42 05 07.2 &  13.87 & 	   & 	   \nl
 J155305.9+445749 & 15 53 05.1 &  +44 57 39.9 &  14.16 & 	   & 	   \nl
 J155532.2+351207 & 15 55 32.7 &  +35 11 54.8 &  13.41 & 	   & 	   \nl
 J155543.2+111114 & 15 55 43.0 &  +11 11 24.1 &  13.81 &  0.360 &   BL\nl
 J155625.4+090311 & 15 56 26.0 &   +9 03 19.2 &  14.47 &  0.042 &  AGN\nl
 J155703.2+635029 & 15 57 03.3 &  +63 50 27.4 &  14.01 &  0.030 &  AGN\nl
 J155721.3+445902 & 15 57 22.6 &  +44 58 54.3 &  14.16 & 	   & 	   \nl
 J155909.5+350144 & 15 59 09.7 &  +35 01 47.3 &  14.14 &  0.031 &  AGN\nl
 J160740.7+254106 & 16 07 40.2 &  +25 41 12.6 &  13.87 & 	   & 	   \nl
 J161004.4+671030 & 16 10 04.0 &  +67 10 25.9 &  13.67 &  0.067 & BL\nl
 J161124.8+585106 & 16 11 24.6 &  +58 51 01.3 &  14.24 &  0.032 &  AGN\nl
 J161301.5+371656 & 16 13 01.7 &  +37 17 15.2 &  14.37 &  0.070 &  AGN\nl
 J161801.9+775230 & 16 17 59.8 &  +77 52 34.5 &  13.55 & 	   & 	   \nl
 J161951.7+405834 & 16 19 51.3 &  +40 58 47.5 &  14.31 &      &     \nl
 J162013.1+400858 & 16 20 12.7 &  +40 09 05.7 &  14.18 & 	   & 	   \nl
 J162409.7+260421 & 16 24 09.3 &  +26 04 31.4 &  14.03 &  0.040 &  AGN\nl
 J162552.9+434654 & 16 25 53.3 &  +43 46 51.9 &  13.70 &   & \nl
 J162903.6+361911 & 16 29 05.3 &  +36 18 58.7 &  14.45 &  0.000 & STAR\nl
 J163056.3+361848 & 16 30 56.1 &  +36 18 48.5 &  14.45 & 	   & 	   \nl
 J165057.5+222653 & 16 50 57.8 &  +22 26 47.8 &  14.29 & 	   & 	   \nl
 J165352.6+394538 & 16 53 52.2 &  +39 45 36.3 &  11.27 &  0.033 & BL\nl
 J165551.7+214559 & 16 55 51.4 &  +21 46 01.2 &  13.79 & 	   & 	   \nl
 J171227.2+355256 & 17 12 28.5 &  +35 53 02.1 &  13.73 &  0.027 &  AGN\nl
 J171959.4+241202 & 17 19 59.7 &  +24 12 07.6 &  14.08 & 	   & 	   \nl
 J172855.8+515654 & 17 28 54.6 &  +51 56 49.2 &  14.21 &      &     \nl
 J173114.5+323250 & 17 31 15.2 &  +32 32 58.4 &  14.05 & 	   & 	   \nl
 J174700.3+683626 & 17 46 59.8 &  +68 36 36.1 &  13.78 & 0.063 & GAL\nl
 J174702.0+493803 & 17 47 03.0 &  +49 38 19.3 &  14.20 & 	   & 	   \nl
 J213740.3+013711 & 21 37 39.9 &   +1 37 16.4 &  13.35 & 	   & 	   \nl
 J222408.1+172903 & 22 24 08.1 &  +17 28 47.4 &  14.42 & 	   & 	   \nl
 J225314.2+040957 & 22 53 11.9 &   +4 10 36.1 &  14.35 &  & \nl
 J225453.7+241449 & 22 54 55.1 &  +24 14 45.5 &  14.18 & 	   & 	   \nl
 J230315.7+085226 & 23 03 15.6 &   +8 52 26.9 &  11.05 &  0.016 &  AGN\nl
 J230706.6+163153 & 23 07 05.3 &  +16 32 27.1 &  13.78 & 	   & 	   \nl
 J231341.0+140113 & 23 13 40.5 &  +14 01 15.6 &  13.86 &  0.041 &  AGN\nl
 J233413.9+073637 & 23 34 13.9 &	+7 37 01.2 &  13.86 & 	   & 	   \nl
 J234106.5+093805 & 23 41 06.6 &	+9 38 09.1 &  14.38 & 	   & 	   \nl
 J234728.8+242743 & 23 47 28.8 &  +24 27 45.8 &  13.65 & 	   & 	   \nl
 J234953.5+242754 & 23 49 53.3 &  +24 27 51.6 &  13.65 & 	   & 	   \nl
 J235122.7+234417 & 23 51 21.5 &  +23 44 24.4 &  14.02 & 	   & 	   \nl
\enddata
\label{tbl-2b}
\end{deluxetable}
\clearpage
\begin{deluxetable}{lrrrrc}
\tablenum{3}
\tablewidth{0pt}
\tablecaption{Other Spectroscopic Identifications}
\tablehead{
\colhead{Name (1RXS)}      & \colhead{R.A.}     &
\colhead{Declination}      & \colhead{$B_{J}$}  &
\colhead{$z$}              & \colhead{Type}}
\startdata
J000011.9+052318 & 00 00 11.80 &  +05 23 17.3 & 16.40 & 0.039 & AGN\nl
J000637.1+434223 & 00 06 36.60 &  +43 42 28.3 & 14.80 & 0.166 & AGN\nl
J000805.6+145027 & 00 08 05.70 &  +14 50 23.3 & 14.50 & 0.043 & AGN\nl
J001409.9+304928 & 00 14 01.00 &  +30 49 24.3 & 18.20 &:0.000 & STAR\nl
J004649.4+152741 & 00 46 50.00 &  +15 27 52.6 & 16.10 &:0.078 & AGN\nl
J005050.6+353645 & 00 50 50.80 &  +35 36 42.2 & 14.60 & 0.056 & AGN\nl
J005346.9+223209 & 00 53 46.20 &  +22 32 22.5 & 15.80 & 0.148 & AGN\nl
J011205.2+224452 & 01 12 05.80 &  +22 44 38.6 & 15.80 & 0.000 & STAR\nl
J020012.5+130317 & 02 00 13.90 &  +13 03 13.1 & 16.20 & 0.000 & STAR\nl
J130710.9+394540 & 13 07 11.50 &  +39 45 33.1 & 15.40 & 0.076 & AGN\nl
J144240.3+262330 & 14 42 40.80 &  +26 23 32.4 & 16.40 &:0.110 & AGN\nl
J151601.5+020055 & 15 16 01.40 &   +2 00 59.8 & 16.80 & 0.105 & AGN\nl
J170320.4+373731 & 17 03 20.10 &  +37 37 23.8 & 16.20 &     * & AGN\nl
J171235.5+245037 & 17 12 35.80 &  +24 50 26.8 & 17.50 &     * & AGN\nl
J221832.8+192527 & 22 18 31.30 &  +19 25 42.8 & 14.40 & 0.000 & STAR\nl
J232841.4+224853 & 23 28 42.90 &  +22 49 43.7 & 15.20 & 0.000 & STAR\nl
J234114.9+142820 & 23 41 15.90 &  +14 28 43.7 & 16.80 & 0.000 & STAR\nl
J235959.1+083355 & 23 59 59.30 &   +8 33 54.1 & 15.40 & 0.083 & AGN\nl
\label{tbl-3}
\enddata
\end{deluxetable}
\clearpage
\begin{deluxetable}{lrrcr}
\tablenum{4}
\tablewidth{0pt}
\tablecaption{The USNO Spectroscopic Follow-up ($0 < \delta < 90$)}
\tablehead{
\colhead{$R_{min}$}           & \colhead{$R_{max}$}   &
\colhead{$AR_{min}$}          & \colhead{$AR_{max}$}  &
\colhead{$AREA$}              }
\startdata
13.50 & 15.40 &  00.00 & 01.00 & 315.250 \nl 
13.50 & 14.55 &  01.00 & 06.00 & 662.000 \nl 
13.50 & 15.40 &  06.00 & 08.00 & 32.750  \nl 
13.50 & 14.40 &  08.00 & 09.00 & 366.750 \nl 
13.50 & 14.40 &  09.00 & 10.00 & 732.750 \nl 
13.50 & 13.90 &  10.00 & 11.00 & 688.000 \nl 
13.50 & 14.40 &  11.00 & 12.00 & 652.750 \nl 
13.50 & 13.90 &  12.00 & 13.00 & 727.566 \nl 
13.50 & 14.60 &  13.00 & 14.00 & 758.042 \nl 
13.50 & 14.60 &  14.00 & 15.00 & 781.193 \nl 
13.50 & 13.70 &  15.00 & 16.00 & 730.482 \nl 
13.50 & 14.60 &  16.00 & 17.00 & 780.750 \nl 
13.50 & 14.60 &  17.00 & 18.00 & 388.000 \nl 
13.50 & 14.90 &  18.00 & 22.00 & 93.500  \nl 
13.50 & 14.50 &  22.00 & 24.00 & 454.750 \nl 
\label{tbl-4}
\enddata
\end{deluxetable}

\begin{deluxetable}{lrrcr}
\tablenum{5}
\tablewidth{0pt}
\tablecaption{The GSC Spectroscopic Follow-up ($0 < \delta < 90$)}
\tablehead{
\colhead{$V_{min}$}           & \colhead{$V_{max}$}    &
\colhead{$AR_{min}$}          & \colhead{$AR_{max}$}   &
\colhead{$AREA$}              }
\startdata
11.00 & 14.00 & 00.00 & 02.00 & 646.250 \nl
11.00 & 14.50 & 02.00 & 04.00 & 298.000 \nl
11.00 & 14.50 & 04.00 & 06.00 &  32.750 \nl
11.00 & 13.50 & 06.00 & 08.00 &  32.750 \nl
13.20 & 14.10 & 08.00 & 09.00 & 366.750 \nl
11.00 & 13.75 & 09.00 & 10.00 & 732.750 \nl
13.96 & 14.10 & 10.00 & 11.00 & 688.000 \nl
13.45 & 13.75 & 11.00 & 12.00 & 652.750 \nl
11.85 & 13.00 & 12.00 & 13.00 & 727.566 \nl
11.00 & 13.35 & 13.00 & 14.00 & 758.042 \nl
11.00 & 13.25 & 14.00 & 15.00 & 781.193 \nl
13.80 & 14.05 & 15.00 & 16.00 & 730.482 \nl
11.00 & 13.75 & 16.00 & 18.00 &1168.750 \nl
11.00 & 14.50 & 18.00 & 20.00 & 000.000 \nl
11.00 & 14.50 & 20.00 & 22.00 & 935.000 \nl
11.00 & 13.50 & 22.00 & 24.00 & 454.750 \nl
\label{tbl-5}
\enddata
\end{deluxetable}
\clearpage

\begin{deluxetable}{cccc}
\tablenum{6}
\tablewidth{0pt}
\tablecaption{The Differential QSO Counts}
\tablehead{
\colhead{ $B$ interval  }       & \colhead{$<B>$}    &
\colhead{ N }                   & \colhead{$\log{\rm surf.den.}$}\\
 & & & $mag^{-1} deg^{-2}$
}
\startdata
12.5-13.5 & 13.0 & ~3 & -3.34 \nl
13.5-14.5 & 14.2 & 19 & -2.69 \nl
14.5-15.5 & 14.8 & 24 & -2.04 \nl
\label{tbl-counts}
\enddata
\end{deluxetable}
\clearpage
\begin{deluxetable}{cccc}
\tablenum{7}
\tablewidth{0pt}
\tablecaption{The QSO Luminosity Function at $0.04 < z < 0.3$}
\tablehead{
\colhead{ $M_B$ interval  }       & \colhead{$<M_B>$}    &
\colhead{ N }                   & \colhead{$\log {\rm LF}$}\\
 & & & $Mpc^{-3} mag^{-1}$
}
\startdata
-21.5 -22.5 & -22.04  & ~3 & -6.05 \nl
-22.5 -23.5 & -23.31  & ~9 & -6.84 \nl
-23.5 -24.5 & -24.10  & 23 & -6.89 \nl
-24.5 -25.5 & -25.10  & 11 & -7.75 \nl
-25.5 -28.5 & -26.72  & ~2 & -9.35 \nl
\label{tbl-LF}
\enddata
\end{deluxetable}
\clearpage
\normalsize

%
%

\clearpage

\begin{figure}
\plotone{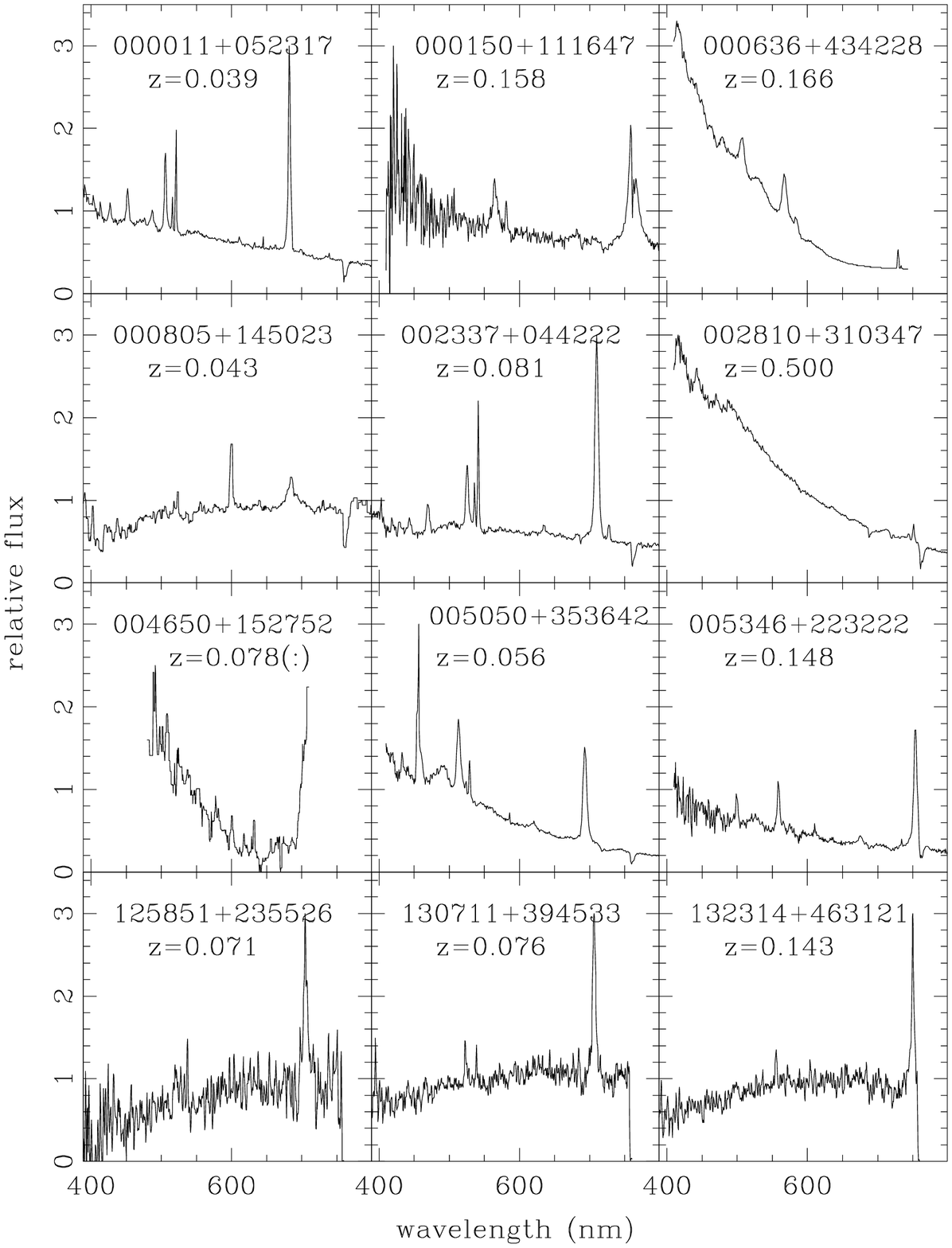}
\caption{a) - The spectra of the AGN confirmed with the follow-up
spectroscopy}
\end{figure}
\begin{figure}
\figurenum{2}
\plotone{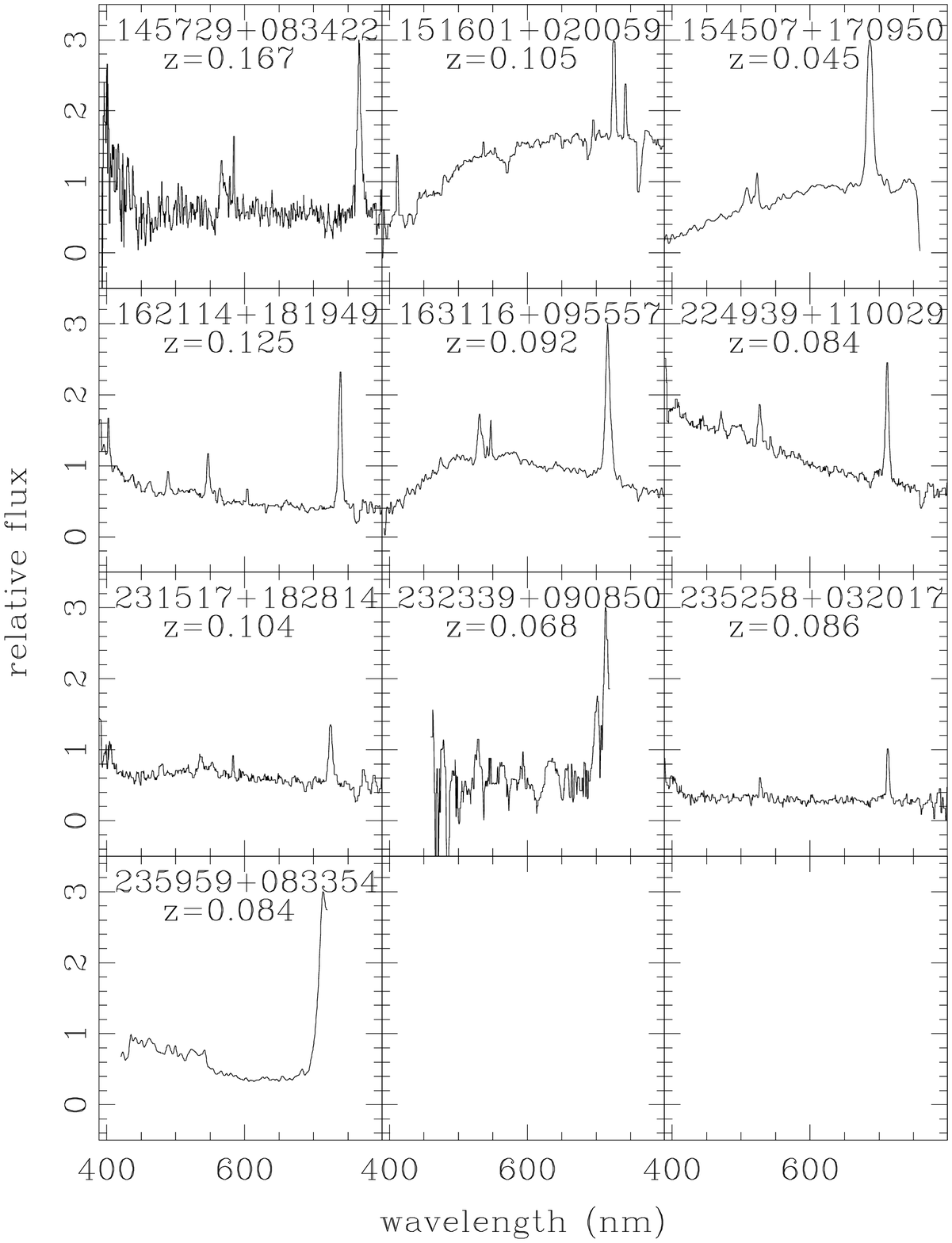}
\caption{b) - The spectra of the AGN confirmed with the follow-up spectroscopy 
\label{fig2}}
\end{figure}
\begin{figure}
\plotone{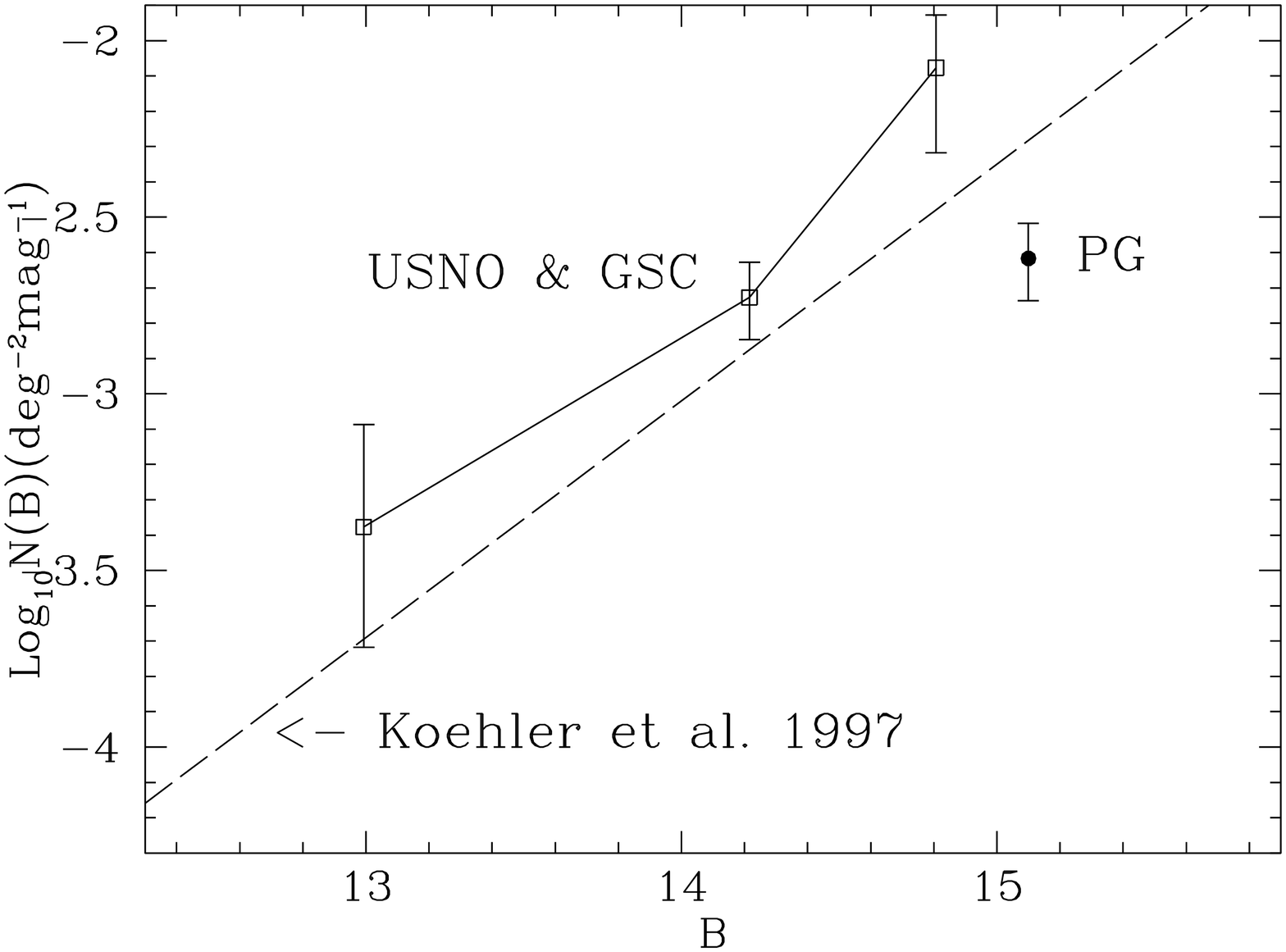}
\caption{The LogN-LogS relation of QSOs.  Open squares refer to the
present sample and are QSOs with $z \ge 0.04$. A correction of -0.037
to the values of Col.~4 in Tab.~6 has been applied to account for the
Bennet bias.  The continuous straight line is the relation found by
K\"ohler et al. (1997) for QSOs with 0.07$\le z\le$2.2. The filled
circle is the point derived from the PG Survey.
\label{fig3}}
\end{figure}
\begin{figure}
\plotone{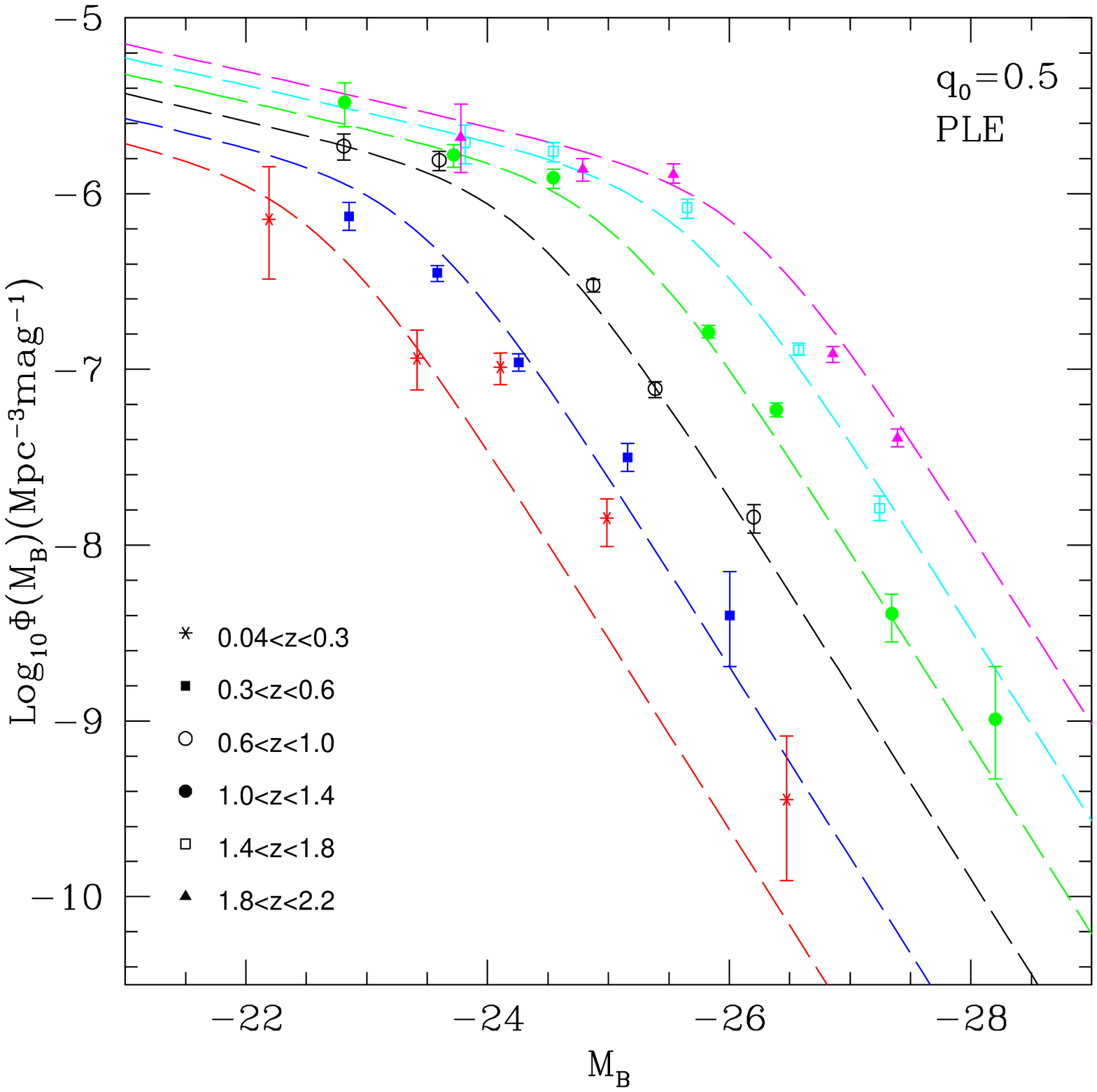}
\caption{a) - The luminosity function of QSOs compared with a
parameterization of Pure Luminosity Evolution (see text). The points
in the range $0.04 < z \le 0.3$ are the result of the present survey, the
remaining data are derived from LC97.}
\end{figure}
\begin{figure}
\figurenum{4}
\plotone{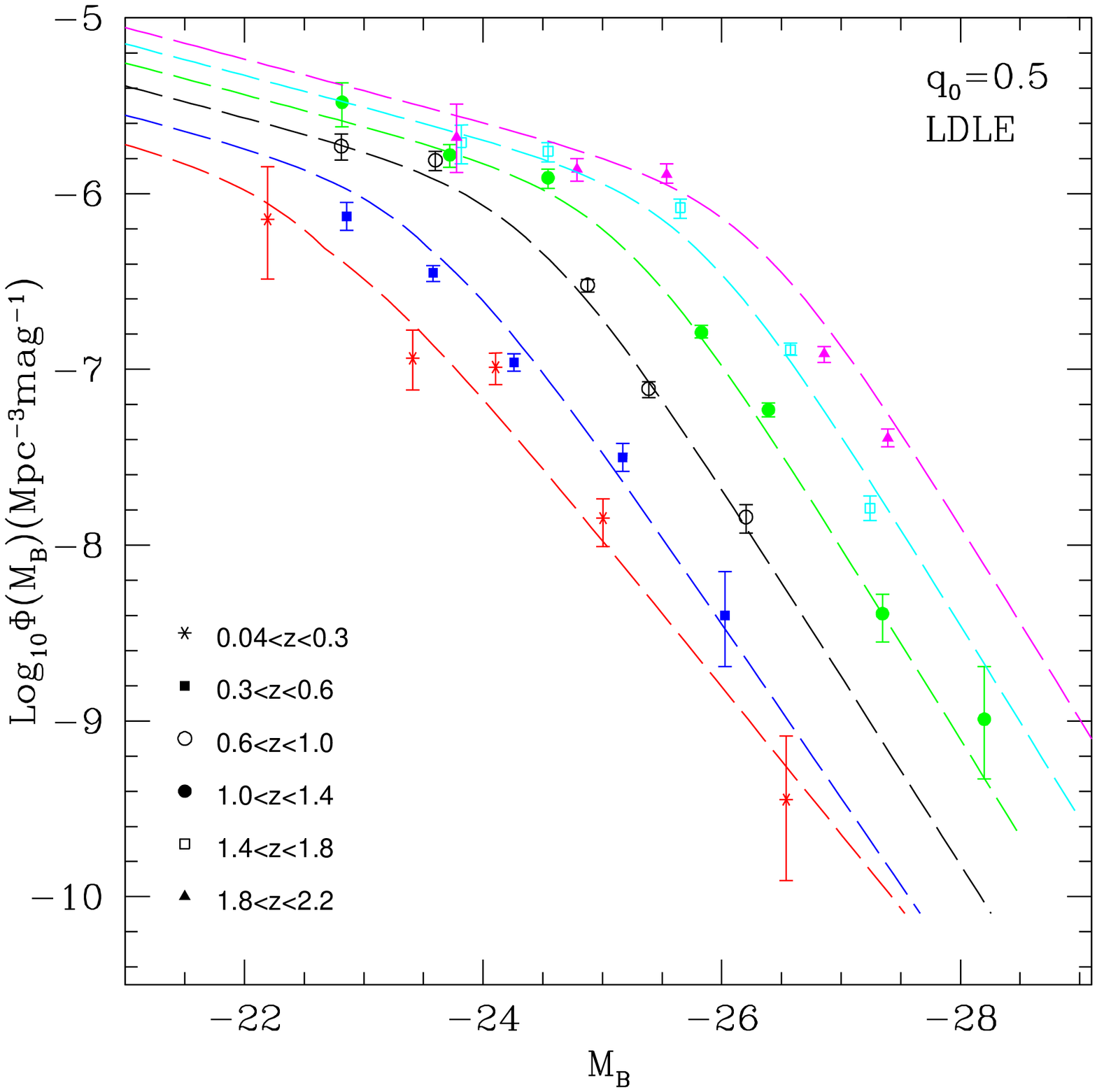}
\caption{b) - The luminosity function of QSOs compared with a
parameterization of Luminosity Dependent
Luminosity Evolution.
\label{fig4}}
\end{figure}
\begin{figure}
\figurenum{5}
\plotone{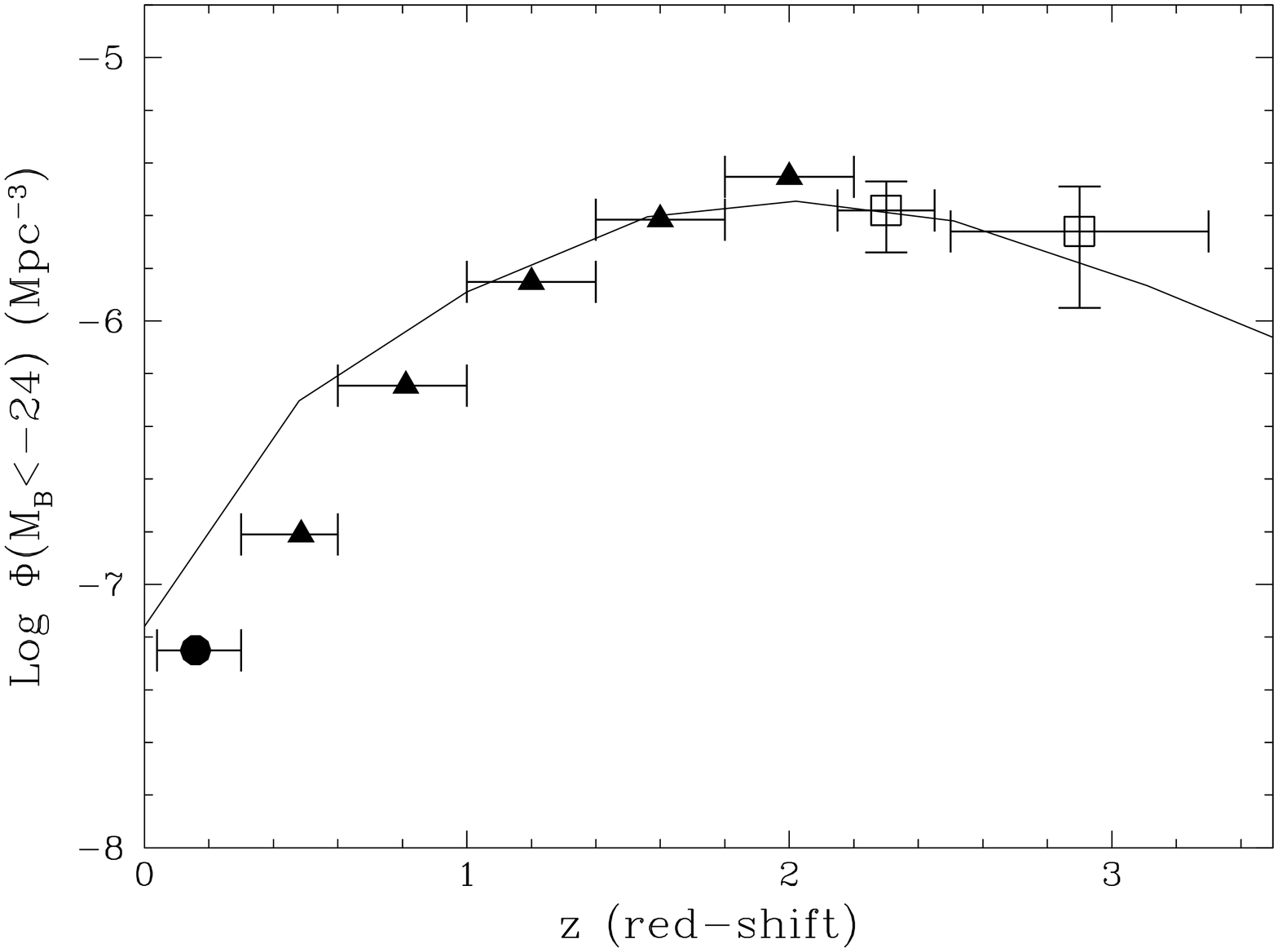}
\caption{The continuous line represents the evolution of the space density of
quasars with $M_B < -24$ predicted by the $\Lambda$CDM model described in the
text (KH99).  
Filled circles are data from the present work, filled triangles are
derived from LC97 and
open squares show data from Hartwick \& Schade (1990).
\label{semi_anal}}
\end{figure}

\begin{thebibliography}{}
\bibitem[Avni \& Bahcall, 1980]{avni:80}
Avni, Y., \& Bahcall, J. N. 1980, ApJ, 235, 694
\bibitem[Banse et al. 1983]{MIDAS} 
Banse, K., Crane, P., Ounnas, C. \& Ponz, D., 1983. In: Proc. of
DECUS, Zurich, p.\ 87
\bibitem[Bessel 1990]{bess90} Bessel, M. S. 1990, \aaps, 83, 357.
\bibitem[Boyle 1992]{boyle92}
Boyle B.J., 1992, in "Texas/ESO-CERN Symposium on Relativistic
Astrophysics, Cosmology and Particle Physics", ed(s) Barrow J.D.,
Mestel L. and Thomas P., Ann. N.Y. Acad. of Sci., 647, 14
\bibitem[Burstein \& Heiles 1982]{bur:hei}
	Burstein, D., Heiles, C. 1982, AJ, 87, 1165
\bibitem[Cattaneo 1999]{cattaneo:99}
Cattaneo, A. astro-ph/9907335 
\bibitem[Cavaliere et al. 1997]{caval97} Cavaliere, A., Perri, M., 
Vittorini, V. 1997, Mem.S.A.It., 68, 27
\bibitem[Cristiani \& Vio (1990)]{cri:vio}
	Cristiani, S., Vio, R. 1990, A\&A, 227, 385
\bibitem[Cristiani et al. 1995]{SC95}
Cristiani, S., La Franca, F., Andreani, P., Gemmo, A., Goldschmidt, P.,
Miller, L., Vio, R., Barbieri, C., Bodini, L., Iovino, A., Lazzarin, M.,
Clowes, R.G., MacGillivray, H., Gouiffes, C., Lissandrini, C., Savage,
A. 1995, A\&AS, 112, 347
\bibitem[Digitized Sky Survey]{http1} The Digitized Sky Survey
{\it http://arch-http.hq.eso.org/dss/dss}
\bibitem[Franceschini et al. 1994]{franceschini94}
Franceschini, A., La Franca, F., Cristiani, S., Martin-Mirones,
J.M. 1994, MNRAS, 269, 683
\bibitem[Gehrels, 1986]{gehrels:86}
Gehrels, N. 1986, \apj, 303, 346
\bibitem[Goldschmidt et al. 1992]{pippa92} Goldschmidt, P., Miller,
L., La Franca, F., Cristiani, S. 1992, MNRAS, 256, 65p 
\bibitem[Goldschmidt and Miller 1998]{pippa98} Goldschmidt, P., Miller,
L. 1998, MNRAS, 293, 107
\bibitem[Haiman and Menou 1999]{HM:99}
Haiman, Z., Menou, K. astro-ph/9810426
\bibitem[Hartwick and Schade 1990]{HS:90} 
Hartwick, F.D.A., Schade, D. 1990, ARA\&A, 28, 437   
\bibitem[Hasinger et al.\ 1998] {hasing98}
Hasinger, G., Burg, R., Giacconi, R., Schmidt, M., Tr\"umper, J.,
Zamorani, G. 1998, A\&A, 329, 482
\bibitem[Hewett and Foltz 1994]{HF94} 
Hewett, P.C., Foltz, C.B. 1994, PASP, 106, 113 
\bibitem{} 
Hewett, P.C., Foltz, C.B., Chaffee, F.H. 1993, \apj, 406, L43 
\bibitem[Kauffmann and Haehnelt]{KH:99}
Kauffmann, G., Haehnelt, M. astro-ph/9906493
\bibitem[K\"ohler et al. 1997]{HES97} 
K\"ohler, T., Groote, D., Reimers, D., Wisotzki, L. 1997, A\&A,
325, 502
\bibitem[La Franca et al, 1995]{lf95}
La Franca, F., Franceschini, A., Cristiani, S., Vio, R.
1995, A\&A, 299, 19
\bibitem[La Franca and Cristiani, 1997]{lf97}
La Franca, F., Cristiani, S. 1997, AJ, 113, 1517
\bibitem[Landolt 1992]{land92} Landolt, A.U. 1992, AJ, 104, 340.
\bibitem[Lasker et al.\ 1988]{lask88} Lasker, B. M., Sturch, C. R.
et al. 1988, \apjs, 68, 1.
\bibitem[Miyaji et al., 1999]{miyaji:99}
Miyaji, T., Hasinger, G., Schmidt, M. astro-ph/9910410
\bibitem[Monaco et al. 1999]{MSD:99}
Monaco, P., Salucci, P., Danese, L. astro-ph/9907095  
\bibitem[Monet et al.\ 1996]{mone96} Monet, D. \aaps, 188, 5404.
\bibitem[Oke, 1990]{stand}
Oke, J.B. 1990, \aj, 99, 1621.
\bibitem[Page \& Carrera, 1999]{page99}
Page, M.J., Carrera, F.J. astro-ph/9909434 
\bibitem[Schmidt \& Green\ 1983]{SeG83} Schmidt, M., Green, R. F. 
1983, ApJ, 269, 352
\bibitem[V\'eron and V\'eron 1998]{veron98}
V\'eron, M.P., V\'eron, P., 1998, A Catalogue of Quasars and Active
Nuclei, 1998, ESO Scientific report No. 18.
\bibitem[Voges 1992]{vog92} Voges, W., 1992 In ESA, Environment
Observation and Climate Modelling Through International Space Projects. Space
Sciences with Particular Emphasis on High-Energy Astrophysics p 9-19
\bibitem[Voges et al.\ 1999]{vog99}
Voges, W., Aschenbach, B., Boller, T.,
Brauninger, H., Briel, U., Burkert, W., Dennerl, K., Englhauser, J.,
Gruber, R., Haberl, F., Hartner, G., Hasinger, G.,
Pfeffermann, E., Pietsch, W., Predehl, P., Rosso, C., Schmidt, J. H. M. M.,
Tr\"umper, J., Zimmermann, H.-U. astro-ph/9909315
\bibitem[Yuan, Siebert and Brinkmann, 1998]{yuan:98}
Yuan, W., Siebert, J., Brinkmann, W. 1998, A\&A, 334, 498
\end{thebibliography}
\end{document}